\documentclass[twocolumn,showpacs,preprintnumbers,amsmath,amssymb]{revtex4}
\newcommand{\Dagcomp}{^{\vphantom{\dagger}}}
\def\frac#1#2{{#1\over#2}}
\def\dfrac#1#2{{\displaystyle{#1\over#2}}}
\def\tfrac#1#2{{\textstyle{#1\over#2}}}
\usepackage{graphicx}% Include figure files
\usepackage{dcolumn}% Align table columns on decimal point
\usepackage{bm}% bold math
\begin{document}

\title{Energy levels and magneto-optical transitions in 
parabolic quantum dots with spin-orbit coupling}
\author{Pekka Pietil\"ainen$^{\ast\dag}$ and Tapash 
Chakraborty$^{\ast\ddag}$}
\affiliation{$^\ast$Department of Physics and Astronomy,
University of Manitoba, Winnipeg, Canada R3T 2N2}
\affiliation{$^\dag$Department of Physical Sciences/Theoretical 
Physics, P.O. Box 3000, FIN-90014 University of Oulu, Finland}
\date{\today}
\begin{abstract}

We report on the electronic properties of few interacting electrons
confined in a parabolic quantum dot based on a theoretical approach developed 
to investigate the influence of Bychkov-Rashba spin-orbit (SO) interaction 
on such a system. We note that the spin-orbit coupling profoundly influences 
the energy spectrum of interacting electrons in a quantum dot. Here we present 
accurate results for the energy levels and optical-absorption spectra for 
parabolic quantum dots containing upto four interacting electrons, in the 
presence of spin-orbit coupling and under the influence of an externally 
applied, perpendicular magnetic field. We have described in detail about a 
very accurate numerical scheme to evaluate these quantities. We have 
evaluated the effects of SO coupling on the Fock-Darwin spectra for 
quantum dots made out of three different semiconductor systems, InAs, InSb, 
and GaAs. The SO coupling on the single-electron spectra manifests itself by 
primarily lifting of the degeneracy at zero magnetic field, rearrangement of
some of the energy levels at small magnetic fields, and level repulsions
at high fields. These are explained as due to mixing of different spinor
states for increasing strength of SO coupling. As a consequence, the
corresponding absorption spectra reveal anticrossing structures in the two
main lines of the spectra. For the interacting many-electron systems we 
observed the appearence of discontinuities, anticrossings, and new modes 
that appear in conjunction with the two main absorption lines. These additional 
features arise entirely due to the SO coupling and are a consequence of level 
crossings and level repulsions in the energy spectra. An intricate interplay 
between the SO coupling and the Zeeman energies are shown to be responsible 
for these new features seen in the energy spectra. Optical absorption spectra 
for all three types of quantum dots studied here show a common feature: new 
modes appear, mostly near the upper main branch of the spectra around 2 tesla 
that become stronger with increasing SO coupling strength. Among the three 
types of systems considered here, optical signatures of the SO interaction 
is found to be the strongest in the absorption spectra of the GaAs quantum dot, 
but only at very large values of the SO coupling strength, and appears to 
be the weakest for the InSb quantum dot. Experimental observation of
these new modes that appear solely due to the presence of SO coupling would 
provide a rare glimpse on the role of SO coupling in nanostructured quantum systems.
\end{abstract}
\maketitle

\section{Introduction}
Impressive developments in nanofabrication technology have made it
now possible to design the quantum dots and coupled quantum dots at
the nanoscale. These systems comprise of a few electrons that are 
quantum confined, for our present purpose, at the semiconductor interface 
to form the zero-dimensional systems. What is more remarkable is that 
the electronic states in these systems can be precisely controlled via 
the external voltages \cite{nanobio}. Magneto-optical studies of 
parabolic quantum dots (described as {\it artificial atoms} \cite{makchak} 
by us in 1990) have been intensely explored for more than a decade 
\cite{makchak,comment,qdbook,merkt,farinfrared} in order to understand its 
unique electronic and optical properties and as promising candidates for 
optoelectronic devices, applications in optical quantum information 
technology, etc. \cite{david,shields}. As a result of these studies, a very good 
theoretical and experimental understanding of the single-electron 
states in the dot has already been achieved. At the most basic level, 
the solution of the Schr\"odinger equation for an electron confined by 
a harmonic potential, $v_c=\frac12m^*\omega_0 r^2$, $\omega_0$ is the 
confinement potential strength, in the presence of an external perpendicular
magnetic field is well known since the beginning of quantum mechanics 
\cite{qdbook,fock}. The eigenvalues in this case are given by
$$E_{nl}=\left(2n+\left|l\right|+1\right)\hbar\Omega-\tfrac12
l\hbar\omega_c$$
where $n=0,1,2,...$ and $l=0,\pm1...,$ are the principal and azimuthal 
quantum numbers respectively, $\Omega^2=\left[\omega_0^2+
\frac14\omega_c^2\right]$, 
and $\omega_c$ is the cyclotron frequency. Dipole-allowed transitions 
among these energy levels will have energies \cite{makchak,qdbook,merkt}
$$\Delta E_{\pm}=\hbar\Omega\pm\tfrac12\hbar\omega_c.$$
This relation has been verified to great accuracy by a variety
of experiments \cite{comment,qdbook,farinfrared}. Interestingly,
however, the observed magnetic field dependent FIR absorption in 
quantum dots containing more than one electron was found to be 
essentially {\it independent} of the number of electrons confined and 
instead was dominated by the above relation for $\Delta E_{\pm}$ 
\cite{merkt}. It was a rather puzzling result because according to 
this, magneto-optics was clearly incapable of providing any relevant 
information about the effect of mutual interactions of the confined 
electrons.  The puzzle was later resolved by Maksym and Chakraborty 
\cite{makchak,comment,qdbook,bert}, who pointed out that for a 
parabolic QD in an external magnetic field, the dipole interaction 
is a function of the center-of-mass (CM) coordinate alone and the 
inter-electron interaction does not play any role. Despite this 
somewhat disappointing performance of a parabolic dot, FIR 
spectroscopy of QDs (parabolic or otherwise) has generated 
enormous interest for over a decade that is yet to subside 
\cite{farinfrared}. In this paper we include another interesting 
element into the problem, the spin-orbit (SO) interaction. We 
demonstrate here that in the presence of spin-orbit coupling 
the magneto-optical transitions show many new and interesting 
features that can be tuned by the SO coupling. Based on these 
results, we propose that magneto-optical transitions are best 
suited to determine optically the unique effect of SO coupling in 
quantum dots described below \cite{condmat}.

Interest on the role of the spin-orbit coupling in nanostructured 
systems is now at its peak, due largely to its relevance to spin 
transport in low-dimensional electron channels \cite{spintro,ohno}. 
The intriguing possibility of tuning the SO field and thereby 
coherently manipulate electron spins in quantum dots has sparked major
activities in the past years \cite{kuan,vam_PRB,vam_JAP,governale,cremers,%
quasi_exact,tsitsishvili,manuel_041,manuel_042,manuel_043,manuel_02,lucignano,%
destefani,debald,koenemann,bellucci,fransson,spinorbit}.
It is hoped that an improved understanding of spin dynamics in the QDs 
might pave the way for future electronic and information processing, 
especially in quantum computing and quantum communication \cite{loss}. 
Spin degree of freedom is perhaps more advantageous than charge because
unlike charge, spin is not coupled to electromagnetic noise and therefore
have much longer coherence time \cite{bandyo}. Improved knowledge
of the influence of spin-orbit coupling in quantum dots is therefore quite
essential in this pursuit. While the majority of experimental efforts has 
focused on magneto-transport measurements \cite{expt}, here we present 
the results for the optical absorption spectra that are experimentally 
observable and could, in principle, provide an important probe of SO 
coupling in few-electron quantum dots.

The spin-orbit interaction in semiconductor heterostructures can be 
caused by an electric field perpendicular to the two-dimensional electron
gas (2DEG). Riding on an electron, this electric field will be {\it felt} 
as an effective magnetic field lying in the plane of the 2DEG, 
perpendicular to the wave vector $k$ of the electron. The effective Zeeman 
interaction of the electron spin with the field lifts the spin degeneracy 
(internal Zeeman effect). This is usually referred to in the literature 
as the Bychkov-Rashba mechanism. This results in an isotropic spin splitting 
energy $\Delta_{\rm SO}$ at $B=0$ proportional to $k$ \cite{rashba}.

Let us consider an electron in the 2DEG moving with a velocity $\vec v$ 
in the presence of an electric field $\vec E$. In the rest frame of
the electron, this transforms (relativistically) into an effective
magnetic field ${\vec B}_{\rm eff}$,
$${\vec B}_{\rm eff}=-\frac1{2c^2}{\vec v}\times {\vec E}$$
where $c$ is the speed of light. The magnetic moment of the electron
will then couple to ${\vec B}_{\rm eff}$. The resulting spin-orbit 
interaction is,
\begin{eqnarray*}
{\cal H}_{\rm SO} &=& {\vec\mu}\cdot{\vec B}_{\rm eff} \\
 &=& -\frac1{2c^2}{\vec \mu}\cdot ({\vec v}\times{\vec E})=\dfrac{e 
\hbar^2}{4m_0^2c^2}\vec\sigma\cdot(\vec k\times\vec E)\\
\end{eqnarray*}
since,
${\vec\mu}=-\dfrac{e\hbar}{2m_0}\vec\sigma$, and $\vec v=\hbar\vec k/m_0$.
It can be rewritten as
\begin{equation}
{\cal H}_{\rm SO} = \dfrac{e \hbar^2}{4m_0^2c^2}\vec\sigma\cdot(
\vec k\times\vec E)=\alpha'\langle E_z\rangle(-{\rm i}\vec\nabla\times
\vec\sigma)_z,
\label{original}
\end{equation}
where the electric field is aligned along the $z$ axis.

Alternatively, a general spin-orbit Hamiltonian that stems directly 
from the quadratic in $v/c$ expansion of the Dirac equation is 
\cite{bethe}
\begin{equation}
{\cal H}_{\rm SO}=\dfrac{e \hbar}{(2m_0c)^2}
{\vec\nabla}V(\vec r)\cdot(\vec\sigma\times{\vec p}).
\label{gradient}
\end{equation}
The electric field associated with $V(\vec r)$
is ${\vec E}(\vec r)={\vec\nabla} V(\vec r)$, and is directed along
the $z$ direction. The spin-orbit interaction Hamiltonian
\begin{eqnarray*}
{\cal H}_{\rm SO}&=& \dfrac{e \hbar^2}{(2m_0c)^2}\langle E_z\rangle
\frac1{\hbar}(\vec\sigma\times{\vec p})\cdot{\hat n}
= \alpha'\langle E_z\rangle\frac1{\hbar}(\vec\sigma
\times{\vec p})\cdot{\hat n} \\
&=& \alpha'\langle E_z\rangle(\vec\sigma\times\vec k)
\cdot{\hat n}, \\
\end{eqnarray*}
where $\alpha'= e (\hbar/2m_0c)^2$ is then identical to that in Eq.~
(\ref{original}). An important point to note here is that a non-vanishing 
gradient in Eq.~(\ref{gradient}) requires that the system must have 
inversion asymmetry. In the present case that arises from the structural 
inversion asymmetry \cite{zawadzki}. The spin-orbit interaction that we 
are here concerned with is therefore described by the Hamiltonian 
\begin{equation}
{\cal H}_{\rm SO}=\alpha (\vec k\times\vec\sigma)_z
={\rm i}\alpha \left(\sigma_y\frac{\partial}{\partial
x}-\sigma_x\frac{\partial}{\partial y}\right),
\end{equation}
where the $z$ axis is chosen perpendicular to the 2DEG (in the 
$xy$-plane), $\alpha$ is the spin-orbit coupling constant, which is 
sample dependent and is proportional to the interface electric field that
confines the electrons in the $x-y$ plane, $\vec{\sigma}=(\sigma_x, 
\sigma_y, \sigma_z)$ denotes the Pauli matrices, and $\vec k$ is the
planar wave vector. This is the Bychkov-Rashba Hamiltonian \cite{rashba}
that has been receiving of late rather widespread attention \cite{expt}. 

The single-electron Hamiltonian for the 2DEG including
the Bychkov-Rashba term has the form
\begin{eqnarray*}
{\cal H}&=&\frac{{\vec p}^2}{2m^*}+\frac{\alpha}{\hbar}
\left(\vec\sigma\times{\vec p}\right)_z \\
&=&-\frac{\hbar^2}{2m^*}{\vec\nabla}^2+{\rm i}\alpha\left(\sigma_y
\frac{\partial}{\partial x}-\sigma_x\frac{\partial}{\partial y}\right)\\
&=&{\left( \begin{array}{lr}
-\dfrac{\hbar^2}{2m^*}{\vec\nabla}^2 & \alpha\nabla^- \\
-\alpha\nabla^+ & -\dfrac{\hbar^2}{2m^*}{\vec\nabla}^2 \\
\end{array} \right)} \\
\end{eqnarray*}
where $\vec{\nabla}^2=\partial^2/\partial x^2+\partial^2/
\partial y^2$ and $\nabla^{\pm}=\partial/\partial
x\pm{\rm i}\partial/\partial y$.

Since the operators ${\hat p}_x$ and ${\hat p}_y$ commute with the 
Hamiltonian, we can search for $\alpha\ne0$ eigenstates of the form
$$\Psi(k_x,k_y)={\rm e}^{{\rm i}k_xx+{\rm i}k_yy}\sum_{\sigma}
C^\sigma\vert\sigma\rangle
={\rm e}^{{\rm i}k_xx+{\rm i}k_yy}\left(\begin{array}{c}
C^+\\C^-\end{array}\right),$$
with $\vert\sigma\rangle=\left(\begin{array}{c}1\\0\end{array}\right)$
(spin up) or $\left(\begin{array}{c}0\\1\end{array}\right)$
(spin down). Solutions of
$${\cal H}\Psi(k_x,k_y)={\cal E}\Psi(k_x,k_y)$$
are readily obtained as
$$\Psi^{\pm}(k_x,k_y)=\frac1{\sqrt2}\left(\begin{array}{c}1\\
\dfrac{\pm k_y\mp{\rm i}k_x}k\end{array}
\right){\rm e}^{{\rm i}k_xx+{\rm i}k_yy}.$$

The energy dispersion then consists of two branches
$${\mathcal E}^{\pm}(k)=\frac{\hbar^2}{2m^*}k^2 \pm \alpha k$$ 
with an energy separation $\Delta_{\rm SO}={\mathcal E}^+-{\mathcal 
E}^-=2\alpha k$ for a given $k$. The spin parts of the wave functions 
$\chi^\pm(k_x,k_y)$ are mutually orthogonal and $\langle\chi^\pm|\sigma_z|
\chi^\pm\rangle=0.$ Therefore in the states $\Psi^\pm$ the spins of the
electrons lie in the $xy$-plane and point in opposite directions. In 
addition, 
$$\langle\chi^\pm|\sigma_x|\chi^\pm\rangle=\frac{2k_y}k, \qquad
\langle\chi^\pm|\sigma_y|\chi^\pm\rangle=-\frac{2k_x}k,$$
i.e., the spins are {\it perpendicular} to the momentum $(k_x,k_y)$. 
Spatial alignment of spins therefore depends on the wave vector 
\cite{spintro,rashba}. The Fermi surface is a pair of concentric circles 
with radii $k_{F,max}$ and $k_{F,min}$. In the present paper, we are dealing 
with systems having rotational symmetry. The formalism in that case is derived in 
detail in Sect.\,II. 

Several experimental groups \cite{expt} investigating the Shubnikov-de Haas 
(SdH) oscillations in a 2DEG confined at the heterojunctions with a narrow-gap 
quantum well (e.g., InGaAs/InAlAs, InAs/GaSb, etc.) have already established 
that lifting of spin degeneracy results from inversion asymmetry of the structure 
which invokes an electric field perpendicular to the layer. Experimentally observed 
values of the SO coupling strength $\alpha$ lie in the range of 5 -- 45 meV.nm 
\cite{expt}. Energy levels of two interacting electrons confined in a parabolic 
quantum dot in an external magnetic field were recently reported by us for this 
range of SO coupling strength \cite{spinorbit}. In the absence of SO coupling, 
electron-electron interaction causes the ground state energy to jump from one 
angular momentum value to another as the magnetic field is increased 
\cite{makchak,qdbook}. The influence of the SO coupling is primarily to move the 
energy level crossings to weaker fields \cite{spinorbit}. 

Our paper is organized as follows: In Sect.~II, we present the essential formalism for
our study of non-interacting electrons in parabolic QDs in the presence of spin-orbit
interaction. The single-electron basis and the dipole matrix elements are derived
here. We also explain why the dipole-allowed transitions are so significantly
influenced by the presence or absence of the SO interaction. The classic
Fock-Darwin spectra for three different quantum dots systems (InAs, InSb, and 
GaAs) with or without the SO coupling are presented and discussed in detail.
The formalism for the many-electron system in a parabolic QD with SO coupling
is presented in Sect.~III. The complexities of introducing the SO coupling,
in particular for interacting electrons, are made clear in Sect.~III. The
task of finding a suitable numerical technique is even more challenging, and
an approach that is appropriate for our purpose is described in the Appendix. 
Numerical results for the energy levels and the optical absorption spectra 
for the three types of QDs containing upto four interacting electrons are 
presented and discussed in Sect.~IV. It should be pointed out that the low-lying
energy levels calculated here for the single- and multi-electron quantum dots
can, in principle, be observed in transport \cite{rolf_review}, or capacitance
spectroscopy \cite{ashoori}. Given the accute interest on the influence 
of SO coupling in nanostructured systems and the resulting intense activities 
on this topic, it is no surprise that many different theoretical techniques 
have been put forward in the literature. To view our work in proper perspective, 
we present a brief review of many of those theoretical papers in Sect.~V. 
We conclude with a brief outlook for future work along this direction in 
Sect.~VI. For a brief account of our earlier work on SO coupling effects in 
parabolic QDs, see Refs.~\cite{condmat,spinorbit}.

\section{Single-electron picture}
The Hamiltonian for an electron in a parabolic confinement and under
external magnetic field is given by
\begin{eqnarray}
{\cal H}_0
&=&
\frac1{2m^*}\left(\vec p -\frac ec\vec A\right)^2+\frac12 m^*\omega_0^2r^2
\nonumber \\ 
&+&\frac {\alpha}{\hbar}\left[\vec\sigma\times\left(\vec p-\frac ec\vec A
\right)\right]_z+\frac12 g\mu_BB\sigma_z.
\label{spham}
\end{eqnarray}
Here $\vec\sigma$ is the vector of Pauli matrices, i.e.
\begin{eqnarray}
\vec\sigma&=&\sigma_x\vec i+\sigma_y\vec j+\sigma_z\vec k 
\nonumber \\
&=&\left(\begin{array}{cc}0&1 \\ 1&0\end{array}\right)\vec i
+\left(\begin{array}{cc}0&-i \\ i&0\end{array}\right)\vec j
+\left(\begin{array}{cc}1&0 \\ 0&-1\end{array}\right)\vec k.
\label{pauli}
\end{eqnarray}
We work in the symmetric gauge and the vector potential corresponding
to the external perpendicular magnetic field is
\begin{equation}
\vec A=\frac B2(-y,x,0).
\label{symmgauge}
\end{equation}
The term $\frac {\alpha}{\hbar}\left[\vec\sigma\times
\left(\vec p - \frac ec\vec A\right)\right]_z$ in the Hamiltonian
is the spin-orbit (SO) coupling due to the inhomogenous potential
confining the electrons to the 2D plane and possible external gate
voltages applied on the top of the dot. The parameter $\alpha$
determines the strength of this coupling and, in case of external
gate voltages its magnitude can be varied. Finally, the last
term $\frac12 g\mu_BB\sigma_z$ is the ordinary Zeeman coupling $g$
being the effective Lande' $g$-factor.

The eigenstates of the single particle problem
\begin{equation}
{\cal H}_0\phi=\varepsilon\phi
\label{sparteq}
\end{equation}
are clearly two-component spinors
\begin{equation}
|\lambda\rangle=\left(\begin{array}{c}\phi^{\uparrow} \\ 
\phi^{\downarrow}\end{array}\right).
\label{spinordef}
\end{equation}
Writing the equation (\ref{sparteq}) in polar coordinates and substituting
a trial wave function of the form
\begin{equation}
\phi=\left(\begin{array}{c}
f^{\uparrow}(r)\,e^{i\ell^{\uparrow}\theta} \\
f^{\downarrow}(r)\,e^{i\ell^{\downarrow}\theta}
\end{array}\right)
\label{trialwf}
\end{equation}
it is easy to see that the quantum numbers $\ell^{\uparrow}$ and
$\ell^{\downarrow}$ must be integers and that they depend on each other
in the way
\begin{equation}
\ell^{\uparrow}=\ell^{\downarrow}-1.
\label{mqncond}
\end{equation}
Hence we need only one quantum number for the angular motion, i.e.
solutions of the single particle equation (\ref{sparteq})
are of the form
\begin{equation}
|\lambda\rangle=|k,\ell\rangle=\left(\begin{array}{c}
f^{\uparrow}_{k,\ell}(r)\,e^{i\ell\theta} \\
f^{\downarrow}_{k,\ell}(r)\,e^{i(\ell+1)\theta}
\end{array}\right).
\label{twocspin}
\end{equation}
Here the quantum number $k$ is associated with radial motion (and
not to be confused with the wave vector described in Sect.\,I).
The form of the spinor (\ref{twocspin}) simply restates the fact
that under SO coupling the good quantum numbers are related
to $\vec L+\vec S$. In our case the conserved quantity is
\begin{equation}
j=\ell^{\uparrow,\downarrow} + s^{\uparrow,\downarrow}_z=\ell+\tfrac12
\label{spartj}
\end{equation}
where $s_z=\pm\frac12$ depending on the component of the spinor, i.e.,
$+\frac12$ for the upper component and $-\frac12$ for the lower one.

In order to find the radial wavefunctions $f^{\uparrow,\downarrow}$
we transform to dimensionless units by setting
\begin{eqnarray*}
\omega_c&=&\frac{eB}{m^*c},\quad 
a^2=\frac{\hbar}{m^*\omega_0
\left(1+{\omega_c^2}/{4\omega_0^2}\right)^{\frac12}}, \\
x&=&\frac{r^2}{a^2}, \quad 
\beta=\frac{m^*\alpha a}{\hbar^2}, \\
b\Dagcomp_R&=&\frac{ea^2B}{\hbar^2}, \quad 
\eta^\pm=1\pm 2b\Dagcomp_R, \\
\nu_\ell^{\uparrow,\downarrow}&=&
\frac{\ell\omega_c}{4\omega_0
\left(1+{\omega_c^2}/{4\omega_0^2}\right)^{\frac12}}
\pm\frac{g\mu_BB}{4\hbar\omega_0
\left(1+{\omega_c^2}/{4\omega_0^2}\right)^{\frac12}}, \\
\varepsilon&=&2\hbar\omega_0
\left(1+\frac{\omega_c^2}{4\omega_0^2}\right)^{\frac12}\nu,
g^{\uparrow,\downarrow}(x)=f^{\uparrow,\downarrow}(r).
\end{eqnarray*}
Substituting these into the Hamiltonian (\ref{spham}) the
radial part of the equation (\ref{sparteq}) takes the form
\begin{eqnarray}
\label{upeq}
x{g^{\uparrow}}'' &+& {g^{\uparrow}}' +\left(\nu-\frac{\ell^2}{4x}
-\frac x4+\nu^{\uparrow}_\ell\right)g^{\uparrow} 
\nonumber \\
&-&\beta x^{1/2}\left({g^{\downarrow}}'+\frac{\ell+1}{2x}g^{\downarrow}
+b\Dagcomp_Rg^{\downarrow}\right)=0 \\
x{g^{\downarrow}}'' &+& {g^{\downarrow}}'+\left(\nu-\frac{(\ell+1)^2}{4x}
-\frac x4+\nu^{\uparrow}_{\ell+1}\right) g^{\downarrow} 
\nonumber \\
&+&\beta x^{1/2}
\left({g^{\uparrow}}'-\frac{\ell}{2x}g^{\uparrow}-
b\Dagcomp_Rg^{\uparrow}\right)=0
\label{downeq}
\end{eqnarray}
of two coupled differential equations. For $\beta=0$ (i.e., $\alpha=0$)
these equations describe radial motions of two independent
two-dimensional harmonic oscillators with eigenvalues
\begin{equation}
\nu_{n\ell}^{\uparrow,\downarrow}
=n+\frac{|\ell|+1}2+\nu^{\uparrow,\downarrow}_\ell
\label{nunm}
\end{equation}
and the eigenfunctions
\begin{equation}
g_{n\ell}=\sqrt{\frac{n!}{(n+|\ell|)!}}\,e^{-x/2}x^{|\ell|/2}L_n^{|\ell|}(x).
\label{gnm}
\end{equation}
Here $L_n^{|\ell|}$ is the associated Laguerre polynomial defined,
for example, by the formula
$$L_n^{|\ell|}(x)=e^xx^{-|\ell|}\frac{d^n}{dx^n}\left(e^{-x}x^{n+|\ell|}\right).$$
Therefore it is logical to seek the solution for Eqs. (\ref{upeq},
\ref{downeq}) in the form of the expansion \cite{kuan}
\begin{equation}
g^{\uparrow,\downarrow}=\sum_{n=0}c^{\uparrow,\downarrow}_{n,\ell}g_{n,\ell}.
\label{gnmexp}
\end{equation}

In spinor language (\ref{trialwf}) this corresponds to the expansion
\begin{eqnarray}
|\lambda\rangle &=&
\left(\begin{array}{c}
\sum_{n=0}c^{\uparrow}_{n,\ell}g_{n,\ell}(x)\,e^{i\ell\theta} \\
\sum_{n=0}c^{\downarrow}_{n,\ell+1}g_{n,\ell+1}(x)\,e^{i(\ell+1)\theta}
\end{array}\right) \nonumber \\
&=&
e^{i\ell\theta}\sum_{n=0}c^{\uparrow}_{n,\ell}
\left(\begin{array}{c} g_{n,\ell}(x) \\ 0 \end{array}\right)
+ \nonumber \\
&&e^{i(\ell+1)\theta}\sum_{n=0}c^{\downarrow}_{n,\ell+1}
\left(\begin{array}{c} 0 \\ g_{n,\ell+1}(x) \end{array}\right).
\label{spdecomp}
\end{eqnarray}
The coefficents $c^{\uparrow,\downarrow}_{n,\ell}$ can be obtained
by minimizing the expectation value $\langle\lambda|{\cal H}_0|\lambda
\rangle$. At this point it is usefull to relabel our spinors. For example, 
for non-negative values of $\ell$ we set
\begin{eqnarray}
u_{2n}&=&e^{i\ell\theta}
\left(\begin{array}{c} g_{n,\ell}(x) \\ 0 \end{array}\right) \\
u_{2n+1}&=&e^{i(\ell+1)\theta}
\left(\begin{array}{c} 0 \\ g_{n,\ell+1}(x) \end{array}\right) \\
z_{2n}&=&c^{\uparrow}_{n,\ell} \label{evenn} \\
z_{2n+1}&=&c^{\downarrow}_{n,\ell+1}.  \label{oddn}
\end{eqnarray}

The minimization of
$$\langle\lambda|{\cal H}_0|\lambda\rangle =
\sum_{n,n'}z_{n'}z_n\langle u_{n'}|{\cal H}_0|u_n\rangle$$
leads now to the diagonalization of the symmetric tridiagonal
matrix ${\cal H}$ with diagonal
\begin{equation}
\mbox{diag}\, {\cal H}=(\nu_{0,\ell}^{\uparrow},\nu_{0,\ell+1}^{\downarrow},
\ldots,\nu_{n,\ell}^{\uparrow},\nu_{n,\ell+1}^{\downarrow},\ldots)
\label{posdiag}
\end{equation}
and subdiagonal
\begin{eqnarray}
\mbox{subdiag}\,{\cal H}=&&\frac{\beta}2\left(0,\sqrt{\ell+1}\eta^+,
\eta^-,\ldots,\sqrt n \eta^-, \right. \nonumber \\
&&\left. \sqrt{n+\ell+1}\eta^+,\sqrt{n+1}\eta^-,\ldots\right).
\label{possubd}
\end{eqnarray}

The diagonalization yields a set of solutions: the set $\{\nu^{(k,\ell)}\}$ 
of eigenvalues and the set $\{z^{(k,\ell)}_n\}$ of eigenvectors indexed by 
the particular solution $k$ and the fixed angular momentum $\ell$. These
are the energies of the spinors and the expansion coefficients 
(\ref{evenn},\ref{oddn}). For negative values of $\ell$ the tridiagonal matrix 
consists of the diagonal
\begin{equation}
\mbox{diag}\, {\cal H}=(\nu_{0,\ell+1}^{\downarrow},\nu_{0,\ell}^{\uparrow},
\ldots,\nu_{n,\ell+1}^{\downarrow},\nu_{n,\ell}^{\uparrow},\ldots)
\label{negdiag}
\end{equation}
and the subdiagonal
\begin{eqnarray}
\mbox{subdiag}\,{\cal H}=&&-\frac{\beta}2(0,\sqrt{|\ell|}\eta^-,\eta^+,\ldots,
\sqrt n \eta^+, \nonumber \\
&&\sqrt{n+|\ell|}\eta^-,\sqrt{n+1}\eta^+,\ldots).
\label{negsubd}
\end{eqnarray}
The spinors thus obtained comprise our single-particle basis
\begin{eqnarray}
{\cal B}_S&=&\big\{|\lambda_i\rangle\big|\,i=0,1,2,\ldots\big\}
\nonumber \\
&=&\big\{|k,\ell\rangle\big|\,k=0,1,2,\ldots;\,
\ell=0,\pm1,\pm2,\ldots\big\}. \nonumber \\
\label{spbas}
\end{eqnarray}

\subsection{Dipole matrix elements}

According to the Fermi golden rule the intensity of absorption in dipole 
approximation is proportional to the square of the matrix element
$$ I=\langle f|\sum_{i=1}^N r_ie^{\pm i\theta_i}|i\rangle $$
when the transition goes from the initial $N$-particle state $|i\rangle$ 
to the final state $|f\rangle$. To evaluate this we need to know the dipole 
matrix elements $d_{\lambda',\lambda}$ between the spinor states 
$|\lambda'\rangle$ and $|\lambda\rangle$, i.e. the elements
\begin{equation}
d_{\lambda',\lambda}
=\langle\lambda'|r\,e^{\pm i\theta}|\lambda\rangle
=\langle k',\ell'|r\,e^{\pm i\theta}|k,\ell\rangle.
\label{dipel}
\end{equation}

For simplicity we consider only the circular polarization $e^{+i\theta}$ (the 
other circular polarization is obatined by reversing the roles of $\lambda$ and 
$\lambda'$). Substituting expansions (\ref{gnmexp}) into the above
expression (\ref{dipel}) we get
\begin{eqnarray}
d_{\lambda',\lambda}&=&a\delta_{\ell',\ell+1}\sum_n\left[
{c'}_n^\uparrow c_n^\uparrow\sqrt{n+\ell+1}-{c'}_{n-1}^\uparrow c_n^\uparrow
\sqrt n\right. \nonumber \\
&+&\left. {c'}_n^\downarrow c_n^\downarrow\sqrt{n+\ell+2}
-{c'}_{n-1}^\downarrow c_n^\downarrow\sqrt n
\right]
\label{posmdip}
\end{eqnarray}
when $\ell\geq0$,
\begin{eqnarray}
d_{\lambda',\lambda}
&=&a\delta_{\ell',0}\sum_n\left[{c'}_n^\uparrow c_n^\uparrow\sqrt{n+1}
-{c'}_{n+1}^\uparrow c_n^\uparrow\sqrt{n+1}\right. \nonumber \\
%&& \left.\mbox{\hspace{4em}}
&+&  \left. {c'}_n^\downarrow c_n^\downarrow\sqrt{n+1}
-{c'}_{n-1}^\downarrow c_n^\downarrow\sqrt n \right]
\label{n1mdip}
\end{eqnarray}
when $\ell=-1$ and
\begin{eqnarray}
d_{\lambda',\lambda}&=&a\delta_{\ell',\ell+1}\sum_n\left[
{c'}_n^\uparrow c_n^\uparrow\sqrt{n-\ell} -{c'}_{n+1}^\uparrow 
c_n^\uparrow\sqrt{n+1}\right. \nonumber \\
%&& \left.\mbox{\hspace{4em}}
&+&\left. {c'}_n^\downarrow c_n^\downarrow\sqrt{n-\ell-1}
-{c'}_{n+1}^\downarrow c_n^\downarrow\sqrt{n+1} \right]
\label{negmdip}
\end{eqnarray}
when $\ell<-1$. The intensity is then obtained from
${\cal I}\propto \vert d_{\lambda_1\lambda_2}\vert^2$ \cite{halonen}.
In all our figures for the absorption spectra, the size of the points is 
proportional to the calculated intensity.

\begin{figure}[t]
\begin{center}
 \includegraphics[angle=-90, width=.48\textwidth]{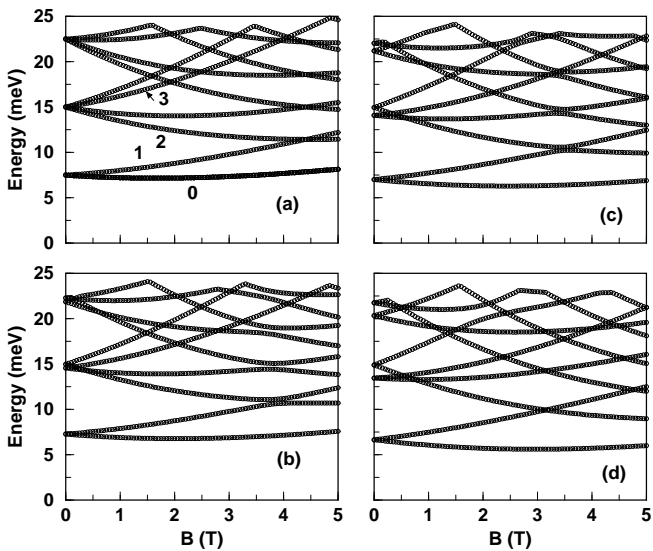}
\protect\caption{
Some of the low-lying energy states for non-interacting electrons confined 
in a InAs quantum dot for various values the SO coupling strength (in meV.nm), 
$\alpha=0$ [a], $\alpha=20$ [b], $\alpha=30$ [c], and $\alpha=40$ [d]. 
The labels in (a) are explained in the text.
}\label{fig:fock1}
\end{center}
\end{figure}

The reason why dipole-allowed transitions in a parabolically
confined quantum dot can be very different in the presence of SO
interaction is explained as follows. When subjected to the radiation
field with amplitude $a$ and polarization $\vec\epsilon$,
the vector potential $\vec A$ in the single particle Hamiltonian
\begin{eqnarray*}
{\cal H}_0&=&\frac1{2m^*}\left(\vec p-\frac ec\vec A\right)^2
+\frac12 m^*\omega_0^2r^2\\
&&+\frac{\alpha}{\hbar}\left[\vec\sigma\times
\left(\vec p-\frac ec\vec A\right)\right]_z+\frac12 g\mu_BB\sigma_z 
\end{eqnarray*}
must be replaced with the potential
$$ \vec A\rightarrow\vec A+ {\vec A}_\omega, 
 \vec A_\omega=\vec\epsilon a \,e^{i\vec k\cdot\vec r-i\omega t}.$$
In the dipole approximation we assume that
$$ A_\omega\approx\vec\epsilon a \,e^{-i\omega t} $$
and correspondingly the Hamiltonian will be \cite{lipparini}
$${\cal H}\approx{\cal H}_0- {\cal H}'\, e^{-i\omega t}, $$
where
$$ {\cal H}'=\frac {ea}{m^*c}\vec\epsilon\cdot\left(
\vec p-\frac ec\vec A\right)
+\frac{\alpha ea}{\hbar c}\left[\vec\sigma
\times\vec\epsilon\right]_z. $$
In a many-body system when $\alpha=0$ the first term
generates the CM density excitations where mutual interactions
play no role. Consequently (in dipole approximation) only
transitions between these modes are possible. When $\alpha$
is different from zero, the second term ($\propto \sigma_x\epsilon_y
-\sigma_y\epsilon_x$) in ${\cal H}'$ can create spin-density
oscillations and interactions have effect on their properties.  
It is to be noted that, in SO coupled systems
the dipole operator still retains its familiar form,
$\hat Q=\dfrac{ea}c\vec\epsilon\cdot\vec r$, as is easily verified 
by evaluating its commutator with the Hamiltonian ${\cal H}_0$
$$ [\hat Q,{\cal H}_0]=i\hbar{\cal H}'. $$
Dipole operator is independent of the electron spin. The 
dipole-allowed optical transitions are always between the same 
spin states, but the angular momenta must differ by unity. 
In the presence of SO coupling, neither the dipole operator 
nor the selection rule changes, but the SO interaction
mixes the neighboring angular momentum values ($l$ and 
$l+1$) as well as the spin and hence the selection rule now 
applies to the total angular momentum $J$ as well. Therefore, 
transitions from other states that are not allowed without 
the SO coupling, are now allowed.

\begin{figure}[t]
\begin{center}
 \includegraphics[angle=-90, width=.48\textwidth]{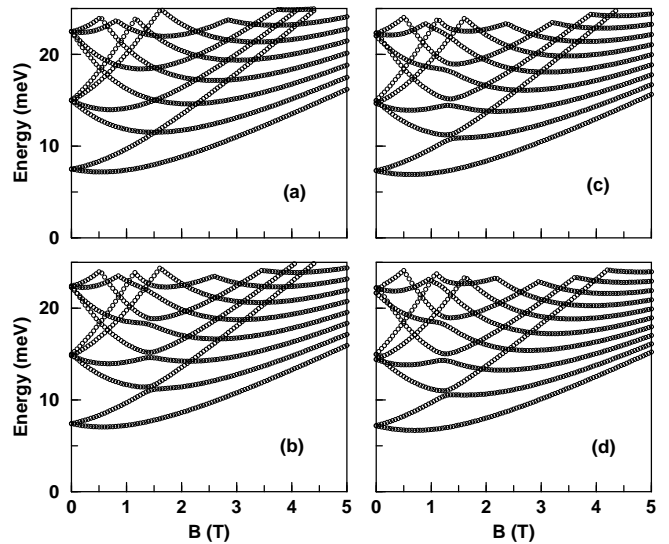}
\protect\caption{
Some of the low-lying energy states for non-interacting electrons confined 
in a InSb quantum dot for various values the SO coupling strength (in meV.nm), 
$\alpha=0$ [a], $\alpha=20$ [b], $\alpha=30$ [c], and $\alpha=40$ [d]. 
}\label{fig:fock2}
\end{center}
\end{figure}

\subsection{Fock-Darwin spectra}

\begin{figure}[t]
\begin{center}
 \includegraphics[angle=-90, width=.48\textwidth]{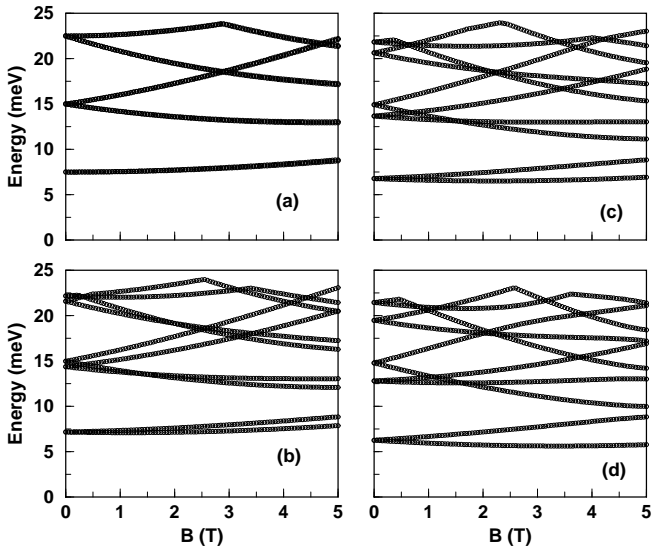}
\protect\caption{
Some of the low-lying energy states for non-interacting electrons confined 
in a GaAs quantum dot for various values the SO coupling strength (in meV.nm), 
$\alpha=0$ [a], $\alpha=20$ [b], $\alpha=30$ [c], and $\alpha=40$ [d]. 
}\label{fig:fock3}
\end{center}
\end{figure}

\begin{figure}[t]
\begin{center}
 \includegraphics[width=.45\textwidth]{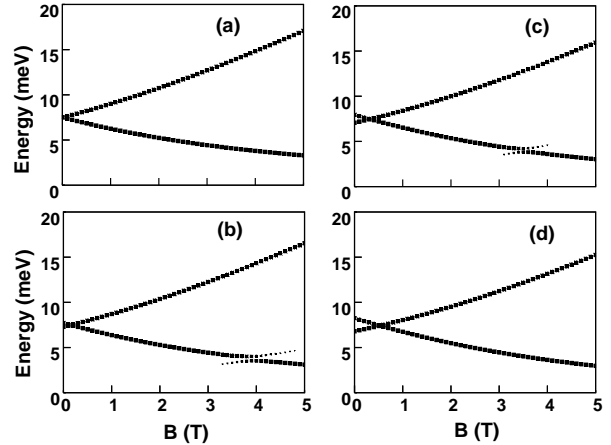}
\protect\caption{
Optical absorption spectra for non-interacting electrons 
confined in a InAs quantum dot 
for various values the SO coupling strength (in meV.nm), $\alpha=0$ 
[a], $\alpha=20$ [b], $\alpha=30$ [c], and $\alpha=40$ [d].
}\label{fig:optics1}
\end{center}
\end{figure}

\begin{figure}[t]
\begin{center}
 \includegraphics[width=.45\textwidth]{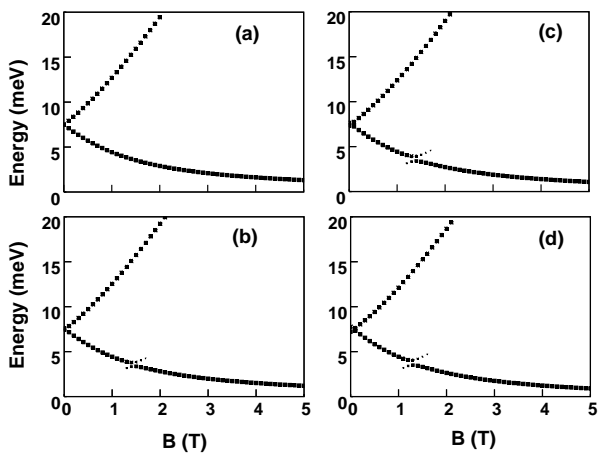}
\protect\caption{
Optical absorption spectra for non-interacting electrons 
confined in a InSb quantum dot 
for various values the SO coupling strength (in meV.nm), $\alpha=0$ 
[a], $\alpha=20$ [b], $\alpha=30$ [c], and $\alpha=40$ [d].
}\label{fig:optics2}
\end{center}
\end{figure}

\begin{figure}[t]
\begin{center}
 \includegraphics[width=.45\textwidth]{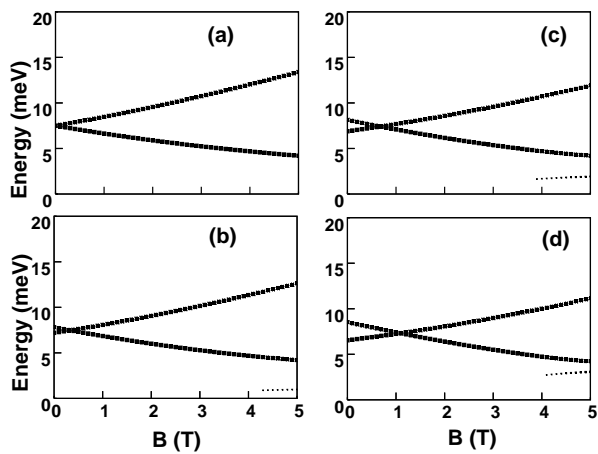}
\protect\caption{
Optical absorption spectra for non-interacting electrons 
confined in a GaAs quantum dot 
for various values the SO coupling strength (in meV.nm), $\alpha=0$ 
[a], $\alpha=20$ [b], $\alpha=30$ [c], and $\alpha=40$ [d].
}\label{fig:optics3}
\end{center}
\end{figure}

In our numerical investigations, we choose InAs, InSb and GaAs quantum 
dots with parameters, $m^*/m_0=0.042, \epsilon=14.6, g=-14$, $m^*/m_0=0.014,
\epsilon=17.88, g=-40$, and $m^*/m_0=0.063, \epsilon=12.9, g=-0.44$, 
respectively. While the InAs quantum structures have been the system of
choice for investigation of spin-related phenomena \cite{expt}, InSb quantum 
dots are interesting for their very high $g$ 
values and a relatively large $\alpha\ (\sim 14$ meV nm) \cite{khodaparast}. 
For the GaAs quantum dots, the observed value of $\alpha$ is $\sim6$ meV.nm 
\cite{koenemann}. In all these systems we consider the confinement potential 
strength to be $\hbar\omega_0=7.5$ meV. Some of the low-lying states of the 
Fock-Darwin spectra of the InAs, InSb and GaAs QDs are shown in 
Figs.~\ref{fig:fock1} -- \ref{fig:fock3} respectively and the corresponding optical 
absorption spectra in these systems 
are presented in Figs.~\ref{fig:optics1} -- \ref{fig:optics3}.

As compared to the Fock-Darwin spectra of quantum dots
without the SO coupling (shown in panels Figs.~\ref{fig:fock1} --
\ref{fig:fock3} [a]) the most outstanding features in the
energy spectra of quantum dots with SO coupling
are the {\it lifting of degeneracy} at zero magnetic
field, the rearrangement of some of the levels at small fields
and level repulsions at higher magnetic fields.
To get some insight into the mechanism causing this kind
of behavior, let us have a closer look, as an example, at
the energy levels involved in the lowest absorption
lines of the InAs dot [curves labelled ``0'' -- ``3'' in 
Fig.~\ref{fig:fock1} (a)].

\begin{table}[h]
\begin{centering}
\begin{tabular}{|r|c|c|}
\hline
&$g<0$&$g>0$ \\
\hline
&& \\
0
&$\displaystyle\left(\begin{array}{c}
\mbox{\rule{1ex}{1.5ex}}\,e^{i0} \\
0e^{i\theta}
\end{array}\right) $
&$\displaystyle\left(\begin{array}{c}
0e^{-i\theta} \\
\mbox{\rule{1ex}{1.5ex}}\,e^{i0}
\end{array}\right)$ \\
&& \\
1
&$\displaystyle\left(\begin{array}{c}
0e^{-i\theta} \\
\mbox{\rule{1ex}{1.5ex}}\,e^{i0}
\end{array}\right)$
&$\displaystyle\left(\begin{array}{c}
\mbox{\rule{1ex}{1.5ex}}\,e^{i0} \\
0e^{i\theta}
\end{array}\right)$ \\
&& \\
2
&$\displaystyle\left(\begin{array}{c}
\mbox{\rule{1ex}{1.5ex}}\,e^{-i\theta} \\
0e^{i0}
\end{array}\right) $
&$\displaystyle\left(\begin{array}{c}
0e^{-2i\theta} \\
\mbox{\rule{1ex}{1.5ex}}\,e^{-i\theta}
\end{array}\right)$ \\
&& \\
3
&$\displaystyle\left(\begin{array}{c}
\mbox{\rule{1ex}{1.5ex}}\,e^{i\theta} \\
0e^{2i\theta}
\end{array}\right) $
&$\displaystyle\left(\begin{array}{c}
0e^{i0} \\
\mbox{\rule{1ex}{1.5ex}}\,e^{i\theta}
\end{array}\right)$ \\
&& \\
\hline
\end{tabular} \\
\end{centering}
\caption{Schematic spinors corresponding to four
Fock-Darwin levels, marked 0 -- 3 in Fig.~\ref{fig:fock1} (a) of an InAs 
dot without SO coupling. The black rectangles stand for non-zero radial 
wave functions.}
\label{spinortable}
\end{table}

In the absence of the SO coupling energies of these levels are
given by the formula (\ref{nunm}). The corresponding spinors (schematic)
are depicted in Table \ref{spinortable}, where the numbers in
the first column refer to the labels in Fig.~\ref{fig:fock1} (a).
The spinors for electrons with negative and positive Lande' $g$-factors
are shown in the middle and third columns respectively. In actual physical systems, 
conventionally only the spinors with $g<0$ are of any interest.

The spinor states of electrons on lines 0 and 1 of Fig.~\ref{fig:fock1} 
(a) differ only by the orientation of the spin: on line 0 the spin is
parallel to the magnetic field ($\|\hat z$) while on line 1 the spin
is antiparallel to the field. Thus the energy difference between these 
states is the Zeemen splitting. The total single-particle angular momenta 
(\ref{spartj}) are correspondingly $j=\pm\frac12$.
Since under the SO coupling $j$ is a good quantum number these two
states will never mix even when the SO coupling is on. When the
coupling strength $\alpha$ increases, the higher lying states
with $j=\frac12$ couple to the state 0 as well as states with $j=-\frac12$
couple with the state 1. In Fig.~\ref{fig:fock1} this shows up
as an increasing splitting of the lines 0 and 1 when going through
the panels from [a] to [d]. It should be noted, however, that
the mixing has very minor effect on the ground state since the
other states with $j=\frac12$ are energetically very far from that.

Turning now our attention to electrons on
lines 1 and 2 we see from the Table \ref{spinortable} (column $g<0$)
that their spinor states both have the same angular momentum
$j=\ell+\frac12=-\frac12$. Consequently the SO interaction can mix these
states. The mixing is particularly pronounced when the states
are nearly degenerate, i.e. in the vicinity of the crossing
point of lines 1 and 2. At moderate coupling strengths this
mixing leads to level repulsions as shown in panels [b] and [c].
When the coupling is very strong, energetically higher states
with $j=-\frac12$ also become important in the mixing, leading to the
imperceptibility of the level repulsion in panel [d].

We also mentioned lifting of degeneracies and rearrangements of 
energy levels as features of the SO coupling under small magnetic 
fields. Since we are interested in the absorption the most important 
states for us are the ones which can be reached from the ground state
($j=\frac12$)  respecting the dipole transition selection rule 
$\Delta j=\pm1$. These are the lowest states with $j=-\frac12$ and 
$j=\frac32$ corresponding to the lines 2 and 3 in Fig.~\ref{fig:fock1} 
(a) and to the spionrs 2 and 3 (with $g<0$) in Table \ref{spinortable}, 
respectively, for vanishing SO coupling. Since as the SO interaction is 
stronger the larger is the (angular) momentum of the electron, the spinor 
3 possessing maximum orbital angular momentum 2 is affected more than the 
spinor 2. Hence, although the energies of both spinors are decreased by 
the SO coupling the effect on the spinor 3 is larger.

Keeping the above discusssions in mind it is now easy to interpret
the features introduced by the SO coupling into the absorption
spectra [Figs.~(\ref{fig:optics1} -- \ref{fig:optics3})]. Firstly, although 
the lower absorption branch consists mainly
of transitions from the state 0 to the state 2 it shows an anticrossing
at moderate coupling strengths. This is a direct consequence of the 
mixing of the spinor states 1 and 2 which results in two spinors, 
both with nonzero upper component. Thus we can see transitions from 
the ground state 0 to both of these. Furthermore, as mentioned earlier
the level repulsion between states 1 and 2 resumes the form of level 
crossing when the SO interaction becomes very strong. This causes the 
anticrossing in the absorption spectra to disappear. In the InAs dot this 
happens already at $\alpha=40$ while in the InSb dot the anticrossing still 
persists. Since the Lande' $g$-factor of GaAs is very small the Zeeman 
split state 1 does not meet the state 2 within the range of the magnetic 
field under consideration. Consequently we see at the lower right corners of 
Fig.~\ref{fig:optics3} (b) -- (d) only the beginnings of the lower branches 
of these anticrossings.

Secondly, the upper absorption branch corresponds mainly to transitions
from the state 0 to the state 3 modified by the SO coupling. The
small magnetic field however makes an exception. As we discussed
above, at small fields the state 3 is energetically lower than the state 2
due to the SO interaction. Thus we get a crossing of spectra at small fields.

\begin{figure}[t]
\begin{center}
 \includegraphics[angle=-90, width=.45\textwidth]{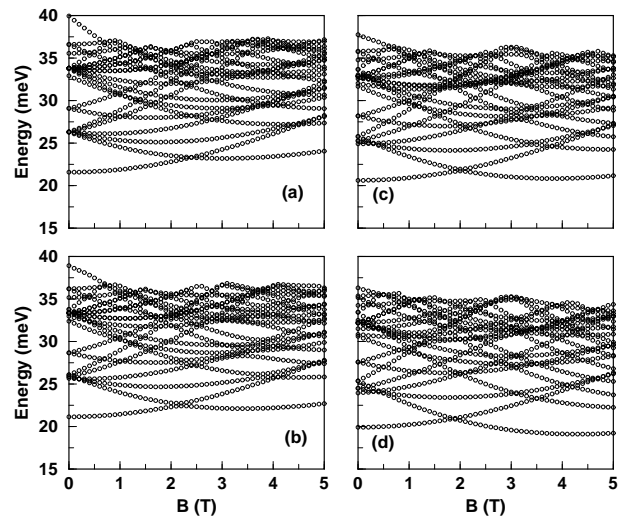}
\protect\caption{
Some of the low-lying energy states for two interacting electrons confined 
in a InAs quantum dot and for various values of the SO coupling 
strength (in meV.nm), $\alpha$ [meV nm]: (a) $\alpha=0$, (b) $\alpha=20$, 
(c) $\alpha=30$, and $\alpha=40$. 
}\label{fig:E2_InAs}
\end{center}
\end{figure}

From Eq.~(\ref{nunm}), the separation between the states 2 and 3, and 
hence also the gap between the absorption line branches is roughly 
proportional to the cyclotron frequency $\omega_c$ which in turn is 
linearly proportional to the magnetic field and inversely proportional 
to the effective mass of the electron. Thus, due to the very small 
effective mass of the electron in an InSb dot the energy of the spinor 3
exceeds the energy of the spinor 2 already at very small magnetic
fields. Consequently also the crossing of absorption lines of an InSb
dot occurs at very small magnetic fields as can be seen in Figs. 
~\ref{fig:optics2} (b)-(d).

\begin{figure}[b]
\begin{center}
 \includegraphics[ width=.45\textwidth]{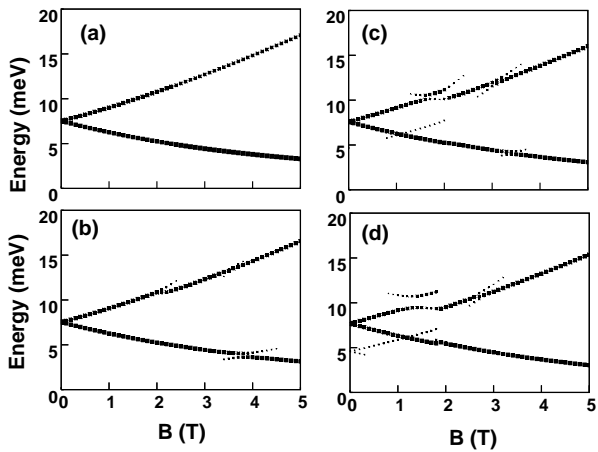}
\protect\caption{
Dipole-allowed transition energies for two interacting electrons confined 
in a InAs quantum dot and for various values of the SO coupling strength (in meV.nm), 
$\alpha$ [meV nm]: (a) $\alpha=0$, (b) $\alpha=20$, (c) $\alpha=30$,
and $\alpha=40$. The size of the points in the figures is proportional to the 
calculated intensity.
}\label{fig:OA2_InAs}
\end{center}
\end{figure}

At this point it may be worth mentioning that the energetics
of the single-electron quantum dot under the influence of the
SO coupling and subjected to an external magnetic field
depends strongly on the sign of the Lande' $g$-factor. This is
contrary to the case without SO coupling where the energy spectrum
is independent on the sign of $g$ although, of course, the orientation
of spin is determined by it. Let us consider, for example, the 
lowest absorption branch in the case $g>0$. From Table \ref{spinortable} 
we can deduce that also in this case the transitions mainly take the 
spinor 0 to the spinor 2. Now, however, there is no spinor level which 
would cross or even come close to the energy of the state 2 and mix with it. 
Consequently the anticrossing described above would not be observable 
in this case.

This concludes our discussion of the energy levels and optical absorption
spectra in a non-interacting QD in a magnetic field and in the presence of 
SO interaction. In what follows, we describe the theory for a interacting 
few-electron parabolic quantum dot.

\section{Many-electron systems}

The basis ${\cal B}_N$ for $N$ interacting electrons in a QD is constructed 
as a direct antisymmetrized
product of single-particle basis ${\cal B}_S$ (\ref{spbas}) of the form
\begin{eqnarray}
{\cal B}_N
&=&{\cal A}\bigotimes_{j=1}^N{\cal B}_S
=\{|\Lambda_i\rangle|\,i=1,2,\ldots\}\nonumber \\
&=&\big\{|\lambda_{i_1};\lambda_{i_2};\ldots;\lambda_{i_N}\rangle
\big|\,i_j=0,1,2,\ldots\big\} \nonumber \\
&=&\big\{|k_1,\ell_1;\ldots;k_N,\ell_N\rangle
\big|\,k_j=0,1,\ldots;\, \nonumber \\
&& \ell_j=0,\pm1,\ldots\big\},
\label{mpbas}
\end{eqnarray}
where ${\cal A}$ stands for the antisymmetrization operator.
It is also understood that the notations such as
$|\lambda_{i_1};\lambda_{i_2};\ldots;\lambda_{i_N}\rangle$
represent the antisymmetrized direct products, i.e.,
\begin{equation}
|\Lambda_q\rangle=
|\lambda_{i_1};\lambda_{i_2};\ldots;\lambda_{i_N}\rangle
={\cal A}\left[
|\lambda_{i_1}\rangle\otimes|\lambda_{i_1}\rangle\otimes
\cdots\otimes|\lambda_{i_1}\rangle\right].
\label{asdirprod}
\end{equation}
Usually it is possible to restrict the size of ${\cal B}_N$ using
the conservation laws. For example, in a rotationally invariant system
the total angular momentum is a good quantum number. Therefore, for example,
we fix it to $J$, and accept into the basis only those states that satisfy
$$\sum_{i=1}^Nj_i=J. $$

\begin{figure}[t]
\begin{center}
 \includegraphics[angle=-90, width=.45\textwidth]{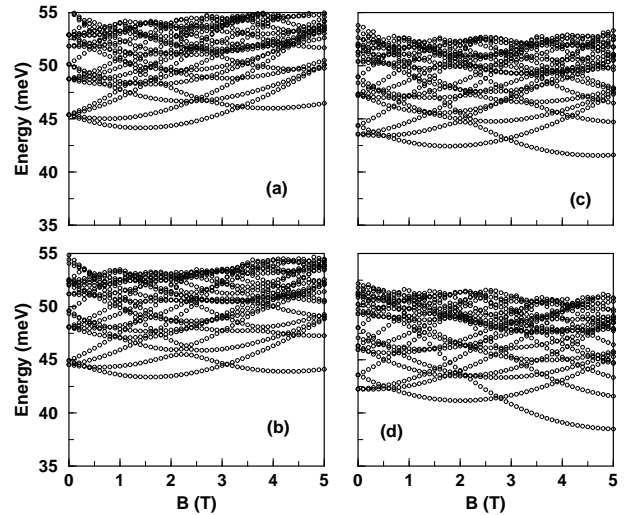}
\protect\caption{Same as in \protect\ref{fig:E2_InAs}
but for a three-electron InAs quantum dot.
}\label{fig:E3_InAs}
\end{center}
\end{figure}

The states of the interacting system are expressed as superposition
of the non-interacting states taken from the basis set (\ref{mpbas})
\begin{equation}
|\Psi\rangle=\sum_{i=1}c_i|\Lambda_i\rangle.
\label{iasuppos}
\end{equation}
To extract the coefficients $c_i$, we again resort to the minimization,
i.e. we minimize the Rayleigh quotient
$$\rho=\frac{\langle\Psi|{\cal H}|\Psi\rangle}{\langle\Psi|\Psi\rangle},$$
where $\cal H$ is the total many-body Hamiltonian.
Again this leads to the diagonalization of the Hamiltonian
matrix with elements
$\langle\Lambda_i\vert{\cal H}\vert\Lambda_j\rangle$. The eigenvectors are
the desired expansion coefficients and the eigenvalues the corresponding
energies of the interacting system. It will be clear from the Appendix 
that both these tasks, construction of the Hamiltonian matrix and its 
diagonalization are numerically quite challenging.

\subsection{Coulomb matrix elements}

\begin{figure}[b]
\begin{center}
 \includegraphics[ width=.45\textwidth]{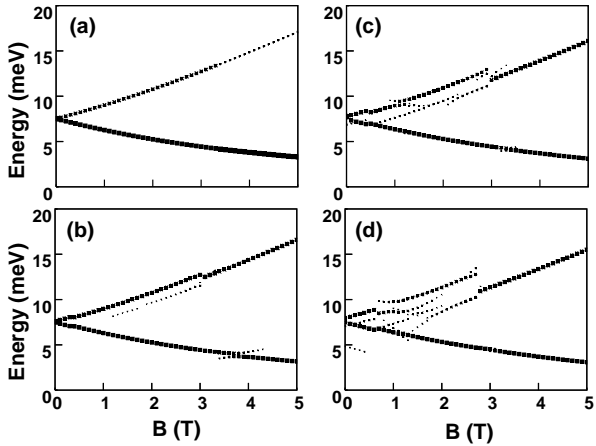}
\protect\caption{Same as in Fig.~\protect\ref{fig:OA2_InAs}, but for
a three-electron InAs quantum dot.
}\label{fig:OA3_InAs}
\end{center}
\end{figure}

We write the total Hamiltonian $\cal H$ as a
sum of the single-particle operators (\ref{spham}) and two-body
operators $V(\vec r, \vec r')$ as
\begin{equation}
{\cal H}=\sum_{i=1}^N{\cal H}_0(\vec r_i)
+\frac12\sum_{i\not=j}^NV(\vec r_i,\vec r_j).
\label{totham}
\end{equation}
Since our basis states $|\Lambda_i\rangle$ are diagonal by construction
in ${\cal H}_0$ we only need to evaluate the matrix elements of
the latter sum in (\ref{totham}). Our many-body states
$|\Lambda_i\rangle$ are expressed in occupation representation
language (\ref{asdirprod}), and therefore it is natural to proceed in
the occupation number space. This means that for the interaction part
we have to evaluate the two-body terms
\begin{equation}
V_{\lambda_1\lambda_2\lambda_3\lambda_4}
=\langle\lambda_1\lambda_2|V|\lambda_3\lambda_4\rangle.
\label{twbterm}
\end{equation}
In our system the mutual interaction between the electrons is taken to be
purely Coulombic, i.e.
\begin{equation}
V(\vec r, \vec r')=\frac{e^2}{\epsilon|\vec r-\vec r'|}
\label{coulpot}
\end{equation}
where $\epsilon$ is the effective dielectric constant of the material.
The interaction operator is thus diagonal in spin. Recalling that
our single-particle states $|\lambda\rangle$ were two-component spinors
(\ref{spinordef}), the two-body term (\ref{twbterm}) consists of sum of 
four terms and is of the form
\begin{equation}
V_{\lambda_1\lambda_2\lambda_3\lambda_4}
=\sum_{\sigma,\sigma'}\int d\vec r\,d\vec r'
{\phi_1^{\sigma}}^\ast(\vec r){\phi_2^{\sigma'}}^\ast(\vec r')
V(\vec r, \vec r')
\phi_3^{\sigma'}(\vec r')\phi_3^{\sigma}(\vec r)
\label{twbtermdec}
\end{equation}
where the summation indices take values $\uparrow$ and $\downarrow$.
Furthermore, since we expressed the spatial components $\phi^\sigma$
as superpositions of functions $g_{n\ell}\,e^{i\ell\theta}$ [Eq.~(\ref{gnmexp})],
$$g^{\uparrow,\downarrow}=\sum_{n=0}c^{\uparrow,\downarrow}_{n,\ell}g_{n,\ell},$$
we are ultimately led to evaluate the Coulomb matrix elements in the
oscillator wavefunction
$$ w_{n\ell}(\vec r) = g_{n\ell}\,e^{i\ell\theta}=\sqrt{\frac{n!}{(n+|\ell|)!}}
\,e^{-x/2}x^{|\ell|/2}L_n^{|\ell|}(x)\,e^{i\ell\theta}$$
basis. These matrix elements can be expressed in terms of finite
sums as \cite{qdbook}
\begin{eqnarray}
&&{\cal A}\Dagcomp_{{\scriptstyle n_1 n_2 n_3 n_4
\atop \scriptstyle \ell_1 \ell_2 \ell_3 \ell_4}}
=\langle w_{n_1\ell_1}w_{n_2\ell_2}|\frac{e^2}{\epsilon|\vec r -\vec r'|}
|w_{n_3\ell_3}w_{n_4\ell_4}\rangle
\nonumber \\
&&=\delta_{\ell_1+\ell_2, \ell_3+\ell_4}
\dfrac{\sqrt2 e^2}{\epsilon a}\left[\dfrac{n_1!}{(n_1+|\ell_1|)!}
\right]^{\frac12}
\nonumber \\
&&\times
\left[\dfrac{n_2!}{(n_2 +|\ell_2|)!}\right]^{\frac12}
\left[\dfrac{n_3!}{(n_3+|\ell_3|)!}\right]^{\frac12}
\left[\dfrac{n_4!}{(n_4+|\ell_4|)!}\right]^{\frac12}
\nonumber \\
&&\times\sum_{\kappa_1=0}^{n_1}\sum_{\kappa_2=0}^{n_2}
\sum_{\kappa_3=0}^{n_3}\sum_{\kappa_4=0}^{n_4}
\left[\kappa_1+\kappa_4+\tfrac12 (|\ell_1|+|\ell_4|-k)\right]!
\nonumber \\
&&\times\left[\kappa_2+\kappa_3+\tfrac12(|\ell_2|+|\ell_3|-k)\right]!
\nonumber \\
&&\times\dfrac{(-1)^{\kappa_1+\kappa_4}}{\kappa_1!\kappa_4!}
\dfrac{(n_1+|\ell_1|)! (n_4+|\ell_4|)!}{(n_1-\kappa_1)!(|\ell_1|+
\kappa_1)!(n_4-\kappa_4)!(|\ell_4|+\kappa_4)!}
\nonumber \\
&&\times\dfrac{(-1)^{\kappa_2+\kappa_3}}{\kappa_2!\kappa_3!}
\dfrac{(n_2 + |\ell_2|)! (n_3+|\ell_3|)!}{(n_2-\kappa_2)!(|\ell_2|+
\kappa_2)!(n_3-\kappa_3)! (|\ell_3|+\kappa_3)!}
\nonumber \\
&&\times\sum_{s=0}^{\kappa_{14}}
\dfrac{\left[\kappa_1+\kappa_4+\frac12(|\ell_1|+|\ell_4|+k)\right]!}{
\left[\kappa_1+\kappa_4+\tfrac12(|\ell_1|+|\ell_4|-k)-s\right]!(k+s)!}
\nonumber \\
&&\times\sum_{t=0}^{\kappa_{23}}
\dfrac{\left[\kappa_2+\kappa_3+\frac12(|\ell_2|+|\ell_3|+k)\right]!}{
\left[\kappa_2+\kappa_3+\frac12(|\ell_2|+|\ell_3|-k)-t\right]!(k+t)!}
\nonumber \\
&&\times\dfrac{(-1)^{s+t}}{s!t!}\cdot\dfrac{\Gamma(k+s+t+
\frac12)}{2^{k+s+t+1}},
\label{coulelsum}
\end{eqnarray}
where
$\kappa_{14}=\kappa_1+\kappa_4+\tfrac12(|\ell_1|+|\ell_4|-k)$,
$\kappa_{23}=\kappa_2+\kappa_3+\frac12(|\ell_2|+|\ell_3|-k)$ and
$k=|\ell_1-\ell_4|=|\ell_2-\ell_3|$.
Numerical techniques to evaluate these two-body terms and to diagonalize the 
resulting Hamiltonian matrix is described in the Appendix.

\begin{figure}[t]
\begin{center}
 \includegraphics[angle=-90, width=.45\textwidth]{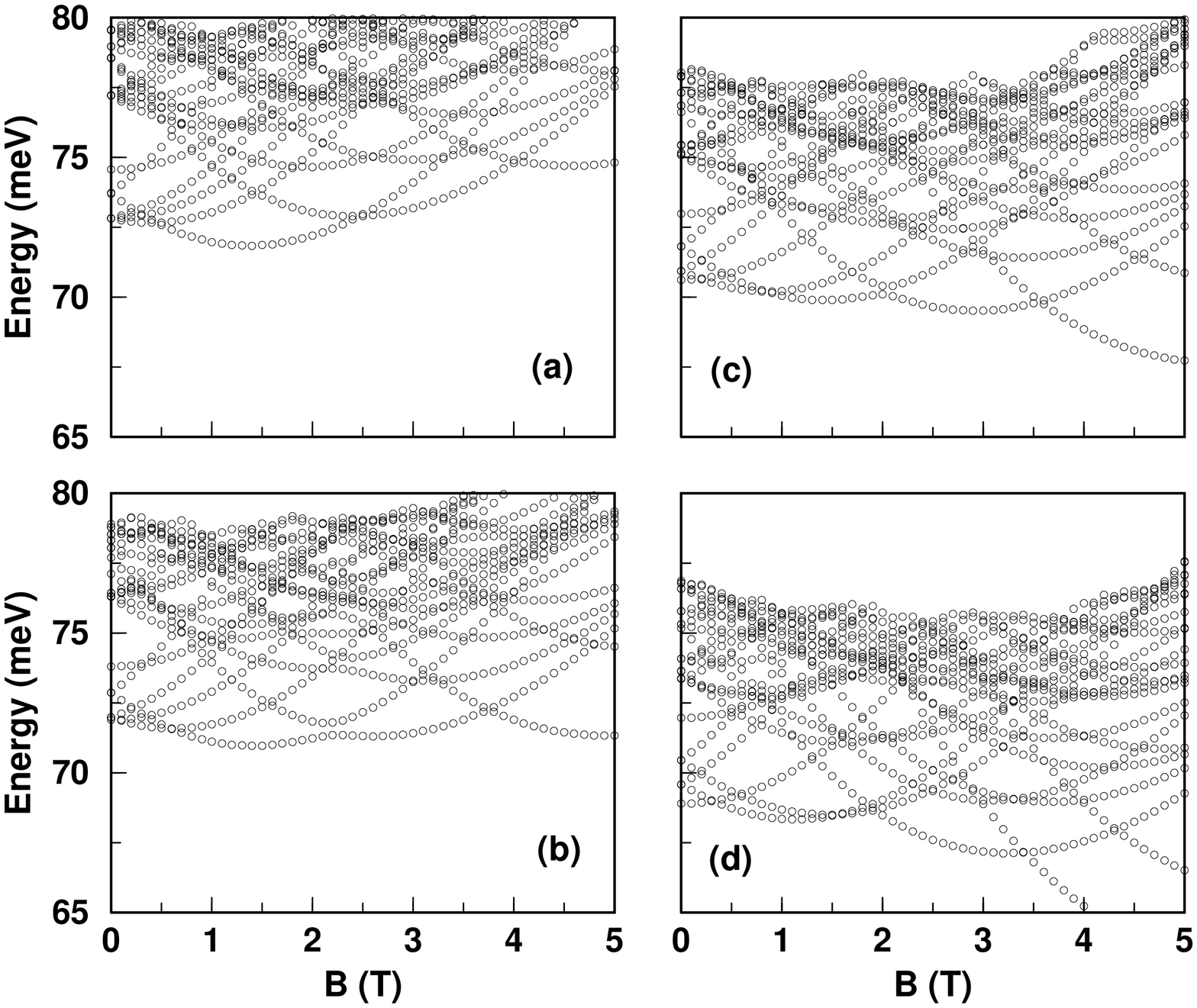}
\protect\caption{Same as in \protect\ref{fig:E2_InAs}
but for a four-electron InAs quantum dot.
}\label{fig:E4_InAs}
\end{center}
\end{figure}

\begin{figure}[t]
\begin{center}
 \includegraphics[ width=.45\textwidth]{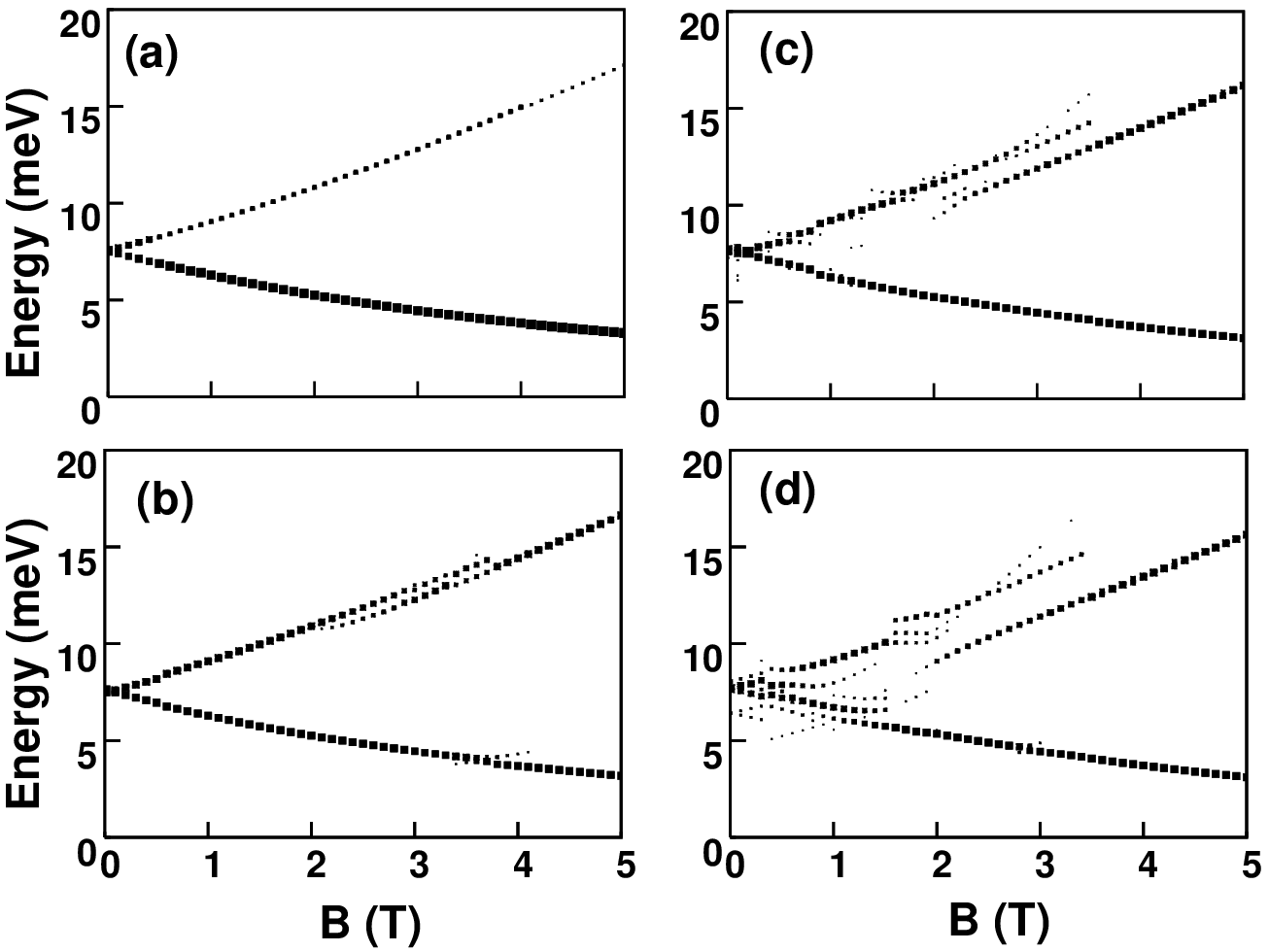}
\protect\caption{Same as in Fig.~\protect\ref{fig:OA2_InAs}, but for
a four-electron InAs quantum dot.
}\label{fig:OA4_InAs}
\end{center}
\end{figure}

\begin{figure}[b]
\begin{center}
 \includegraphics[angle=-90, width=.45\textwidth]{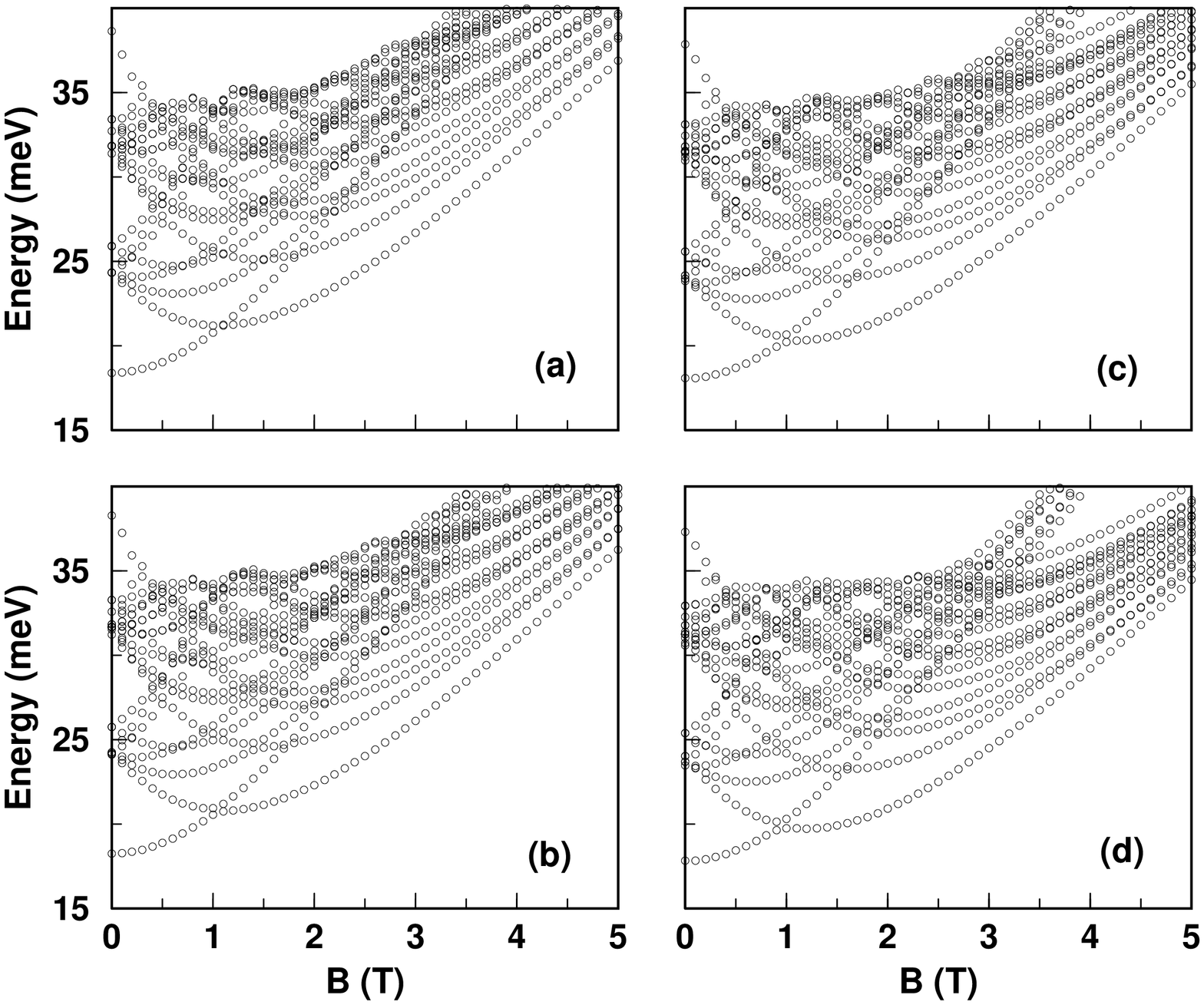}
\protect\caption{Same as in \protect\ref{fig:E2_InAs}
but for a two-electron InSb quantum dot.
}\label{fig:E2_InSb}
\end{center}
\end{figure}

\begin{figure}[t]
\begin{center}
 \includegraphics[ width=.45\textwidth]{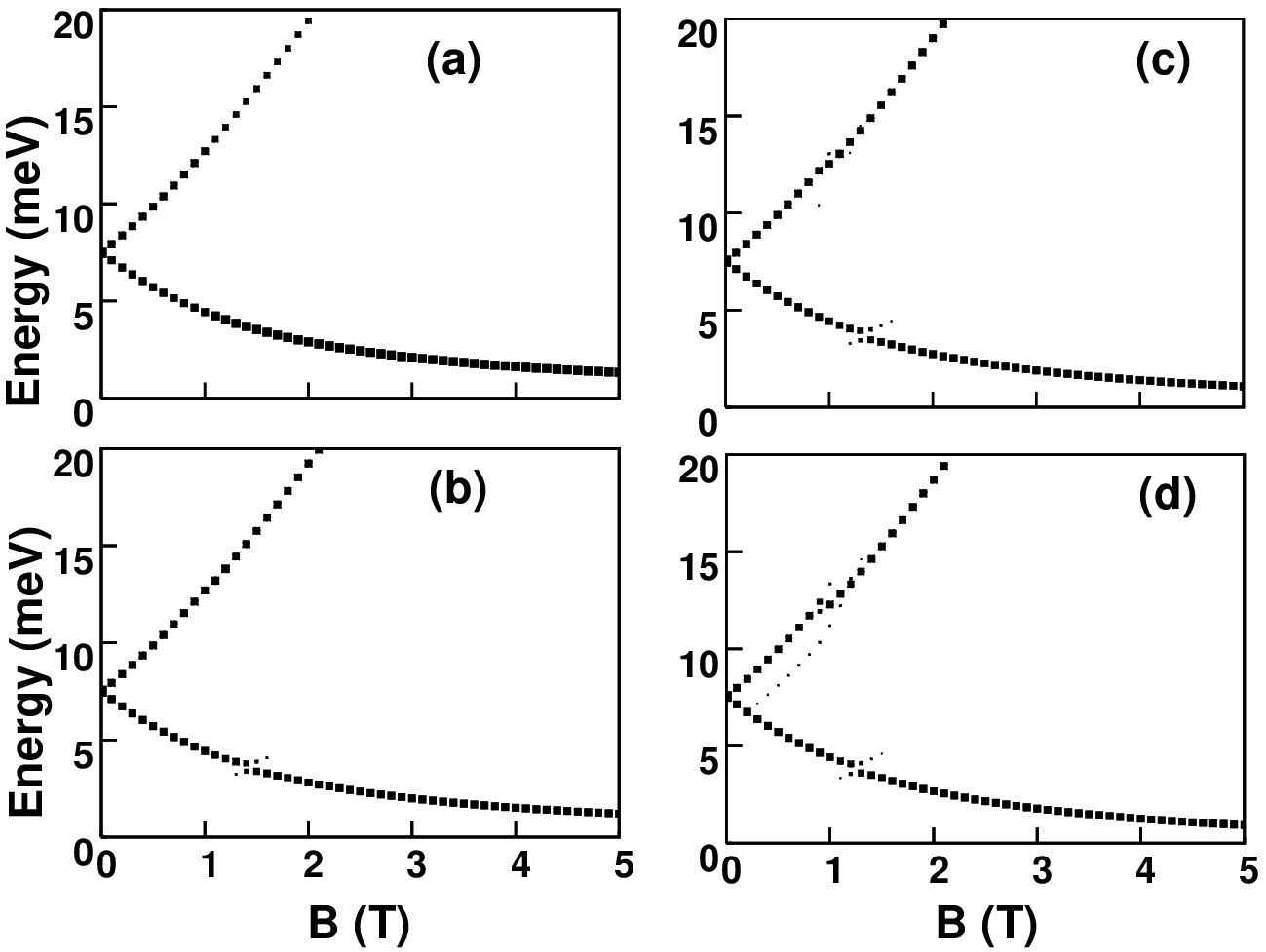}
\protect\caption{Same as in Fig.~\protect\ref{fig:OA2_InAs}, but for
a two-electron InSb quantum dot.
}\label{fig:OA2_InSb}
\end{center}
\end{figure}

\begin{figure}[b]
\begin{center}
 \includegraphics[angle=-90, width=.45\textwidth]{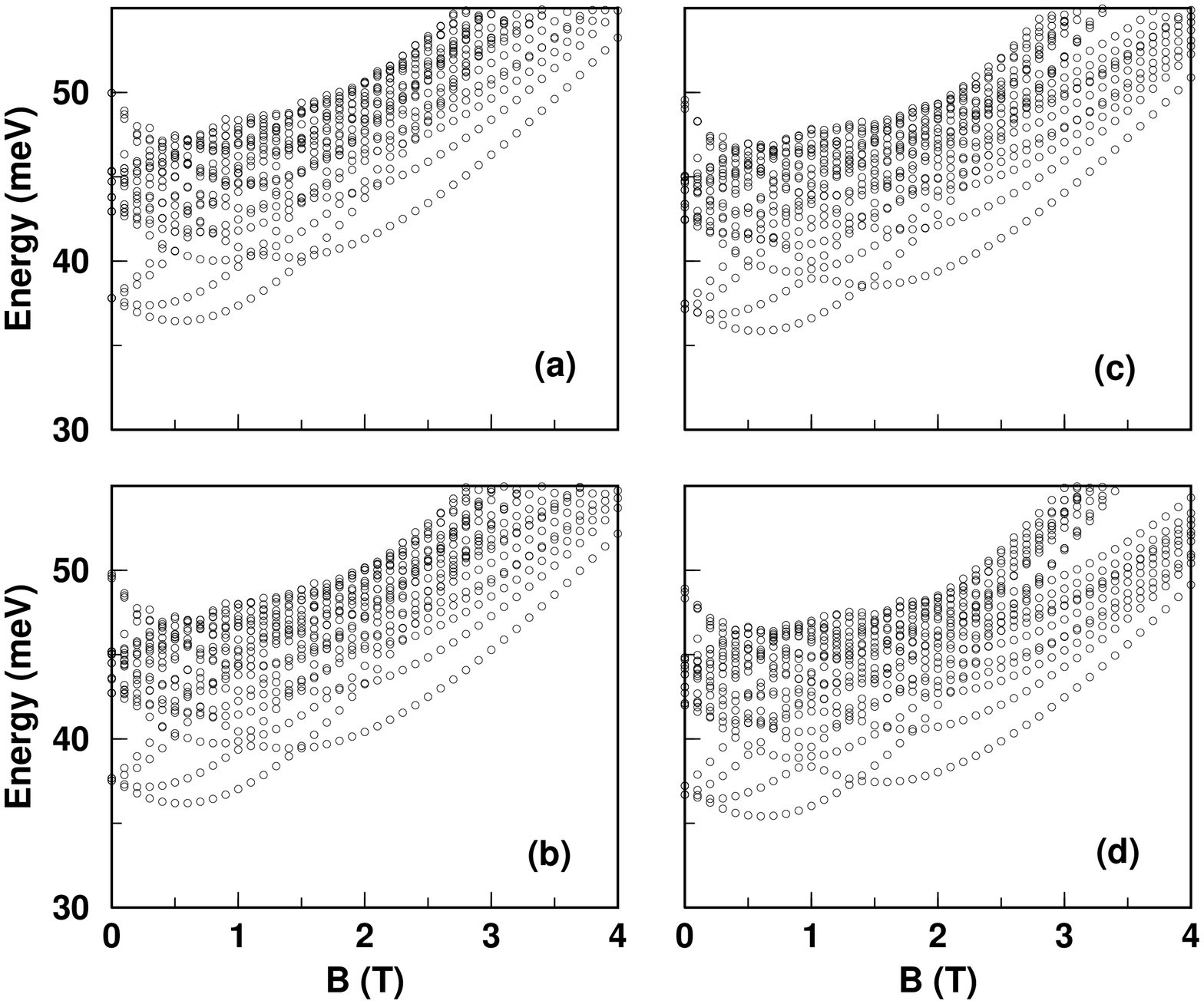}
\protect\caption{Same as in \protect\ref{fig:E2_InAs}
but for a three-electron InSb quantum dot.
}\label{fig:E3_InSb}
\end{center}
\end{figure}

\section{Results and discussions}

For numerical evaluation of the energy spectra and the optical absorption 
spectrum for QDs with a few interacting electrons, we have considered the 
InAs, InSb and GaAs quantum dots. Parameters of these systems are already
given in Sect.\,II.  The energy spectra and the optical absorption spectra
for these systems are described in the subsections below.

\begin{figure}[t]
\begin{center}
 \includegraphics[width=.45\textwidth]{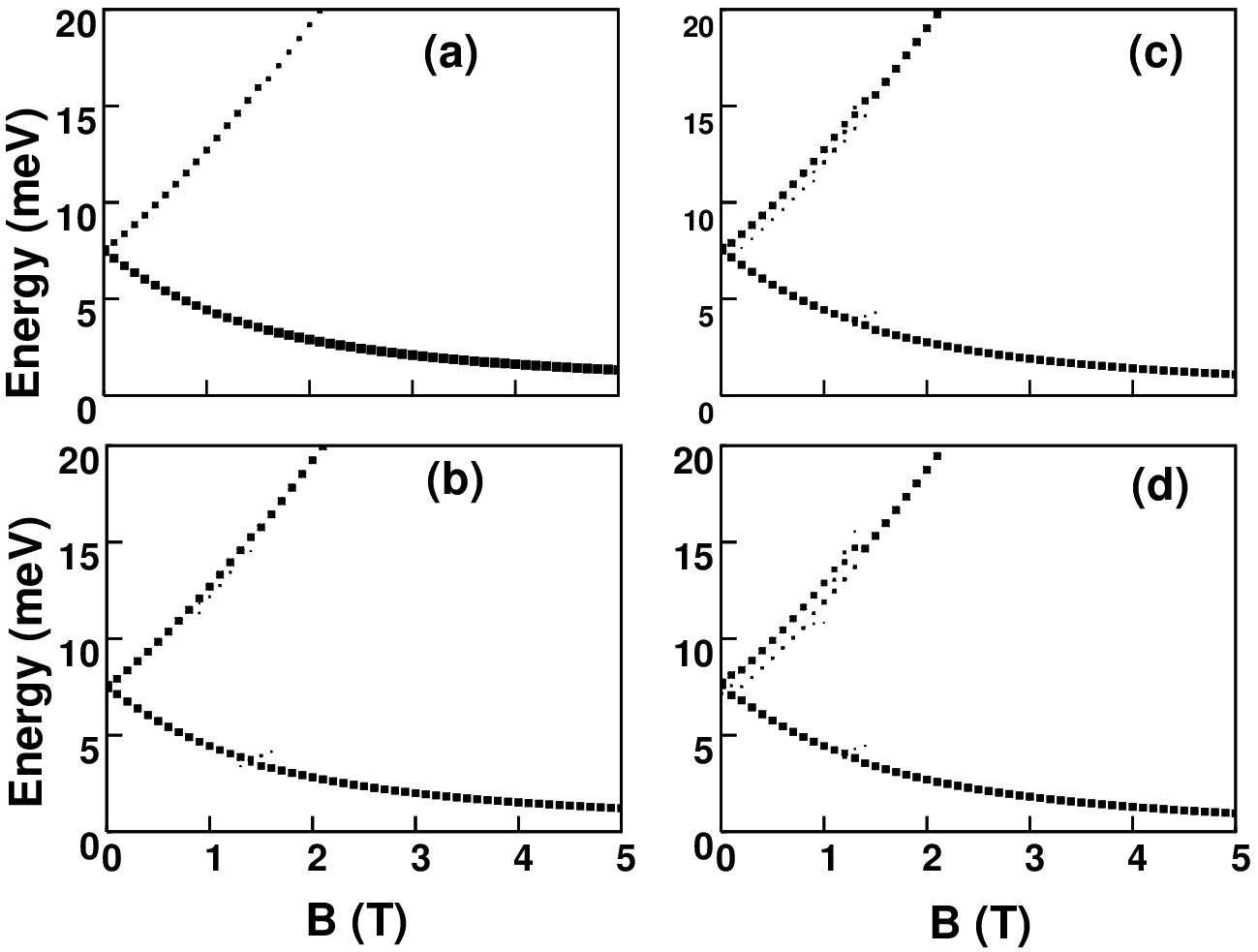}
\protect\caption{Same as in Fig.~\protect\ref{fig:OA2_InAs}, but for
a three-electron InSb quantum dot.
}\label{fig:OA3_InSb}
\end{center}
\end{figure}

\subsection{InAs quantum dots}

Our numerical results for energy spectra and absorption spectra 
(dipole-allowed) for 2 -- 4 electrons are presented in 
Figs.~\ref{fig:E2_InAs}-\ref{fig:OA4_InAs}, and for various values of 
the SO coupling strength (in meV.nm), $\alpha$. As in the case of the non-interacting 
electron system, we have considered the following parameters for the 
InAs quantum dot: $m^*/m_0=0.042, \epsilon=14.6, g=-14$ and 
$\hbar\omega_0=7.5$ meV.

\begin{figure}[b]
\begin{center}
 \includegraphics[angle=-90, width=.45\textwidth]{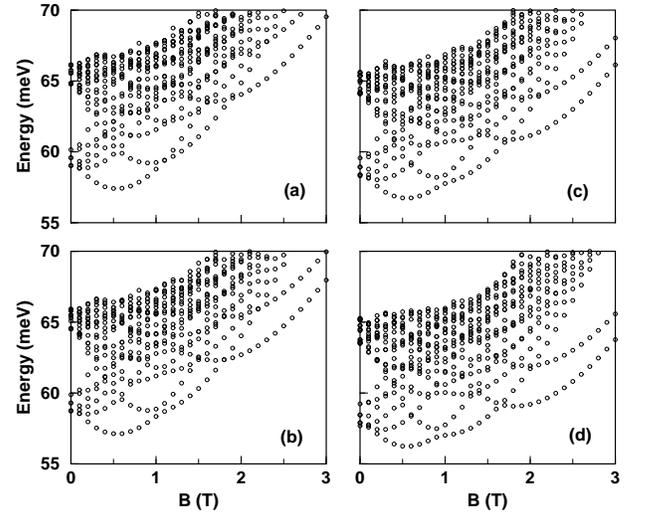}
\protect\caption{Same as in \protect\ref{fig:E2_InAs}
but for a four-electron InSb quantum dot.
}\label{fig:E4_InSb}
\end{center}
\end{figure}

A striking feature visible in the absorption spectra
(Figs.~\ref{fig:OA2_InAs}, \ref{fig:OA3_InAs}, and \ref{fig:OA4_InAs}) is
the appearance of discontinuities, anticrossings and new modes in addition
to the two main ($\alpha=0$) absorption lines. These optical signatures of
the SO interaction are consequences of the multitude of level crossings
and level repulsions that occur in the energy spectra (Figs.~\ref{fig:E2_InAs},
\ref{fig:E3_InAs}, and \ref{fig:E4_InAs}). The latter
ones can be attributed to an interplay between the SO and Zeeman
couplings. In order to understand their origin, let us first examine
the case of the two-electron system
(Figs.~\ref{fig:E2_InAs},\ref{fig:OA2_InAs}).
In our spinor notation the main contribution to the ground state at zero
magnetic field comes from the two-electron state
$|\lambda_{\ell_1},\lambda_{\ell_2}\rangle=|\lambda_0,\lambda_{-1}\rangle$,
where $|\lambda_{\ell_1}\rangle$ is a spinor with $j_1=\ell_1+1/2=1/2$,
$d_n^{\lambda_1}=0$, and $|\lambda_{\ell_2}\rangle$
a spinor with $j_2=-1/2$ and $u_n^{\lambda_2}=0$,
i.e., both electrons have zero orbital angular momenta with opposite spins
(corresponding to the spinors 0 and 1 in Table \ref{spinortable}
with $J=j_1+j_2=0$). When we increase the magnetic field
the spin triplet configuration will become, due to the interaction,
energetically more favorable. If the Lande' $g$-factor is negative then
the electrons would like to occupy states with orbital angular momenta
0 and $-1$ with both spins up (i.e., states 0 and 2 of Table \ref{spinortable}).
In the spinor picture this means that $|\lambda_{\ell_2}\rangle$ still has $\ell_2=-1$ 
($J=0$) but now $u_n^{\lambda_2}\not=0$ and $d_n^{\lambda_2}=0$.
The SO interaction mixes these two configurations which results in
a level repulsion. On the other hand, when the strength of the
SO coupling is further increased, the relative significance of the Zeeman
contribution to ${\cal H}_0$ decreases. The energy shifts to states with
$J\not=0$ will then become energetically feasible and we again have
crossings of levels. For increasing number of electrons in the dot, the
energy spectra is more dense and exhibit additional level crossings
(Figs.~\ref{fig:E3_InAs} -- \ref{fig:OA4_InAs}). As a consequence, the ground 
state angular momentum also changes more frequently as compared to that of 
the two-electron case. It should be pointed out that in many-electron dots these 
level crossings and repulsion are to be attributed, at least partly
to the mutual Coulomb interactions. The level crossings/repulsions
we saw earlier [e.g. levels 1 and 2 in Fig.~\ref{fig:fock1} (a)]
in the single-particle particle picture are due to the Zeeman splitting whereas
in interacting systems crossings occur even in the limit of vanishing
Zeeman coupling. In the present InAs dot the Coulomb interaction brings,
for example the singlet-triplet transition to much lower magnetic field
($B\approx 2$T) as compared to the field required in a noninteracting system 
($B\approx 4$T).

\begin{figure}[t]
\begin{center}
 \includegraphics[width=.45\textwidth]{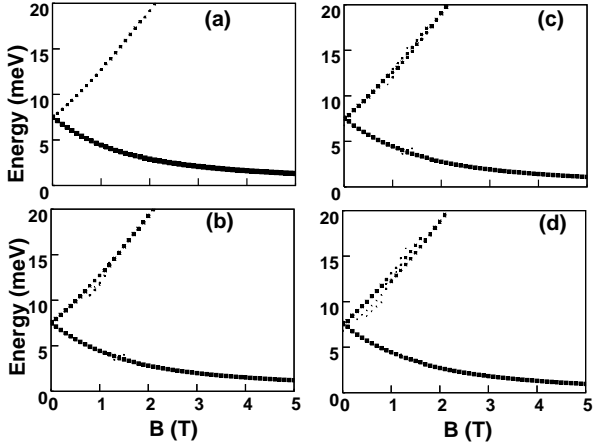}
\protect\caption{Same as in Fig.~\protect\ref{fig:OA2_InAs}, but for
a four-electron InSb quantum dot.
}\label{fig:OA4_InSb}
\end{center}
\end{figure}

At moderate SO coupling strengths the absorption spectra do not
essentially differ from the single-particle spectrum.  But when the coupling
strength increases the deviation from the pure parabolic confinement also
increases which in turn implies that the lowest final states
of dipole allowed transitions are not any more achievable by adding
$\hbar\Omega\pm\frac12\hbar\omega_c$ to the initial state energies.
In particular, this results in discontinuities and anticrossing
behaviors as well as appearence of new modes. As an illustration,
let us consider the absorptions that at a magnetic field of
$B=1$T take the two-electron system from the ground state to excited
states. In the absence of the SO coupling the ground state is a
spin-singlet state $S =0$ with total angular momentum $J=0$.
According to the dipole selection rules absorptions cause transitions to
states $J=\pm1$ and $S=0$ with energies $\Delta E_\pm$ above
the ground state. In Fig.~\ref{fig:OA2_InAs} (d), we note that in addition
to the two main lines there are now two additional lines (at around $B=1$ T)
of appreciable intensity at the SO coupling strength $\alpha=40$ (mev.nm).
Further analysis reveals that the ground states still have $J=0$ and that
the expectation value of the spin $z$-component is $\langle\sigma_z\rangle=0$.
The excited states also have $J=\pm1$, as before. However, the final
spin states can no longer be classified as singlets: the expectation
values $\langle\sigma_z\rangle$ vary between $-0.03$ and 0.39.
When the number of electrons increases the number of these additional
modes also increases but at the same time the relative intensities
decrease (at each $B$ we have normalized the total intensity to unity).
On the other hand, the discontinuities as consequences of deviations
from a parabolic confinement become more pronounced
(Figs.~\ref{fig:E3_InAs} -- \ref{fig:OA4_InAs}). This is because there
are higher angular momenta involved in the dipole transitions. As a consequence 
of this the upper absorption branch now exhibits a rich structure while 
in the single-particle picture it is practically featureless.

\begin{figure}[t]
\begin{center}
 \includegraphics[angle=-90, width=.45\textwidth]{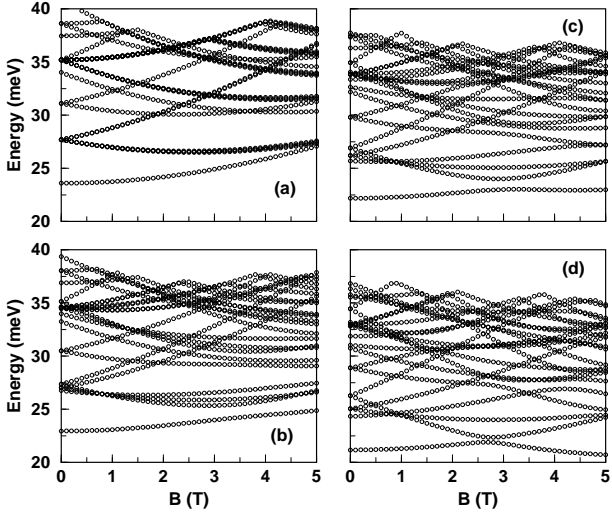}
\protect\caption{Same as in \protect\ref{fig:E2_InAs}
but for a two-electron GaAs quantum dot.
}\label{fig:E2_GaAs}
\end{center}
\end{figure}

\subsection{InSb quantum dots}

As mentioned above, in addition to the InAs quantum dots, investigation 
of InSb quantum dots are also thought to be interesting, particularly 
in the context of SO coupling effects due to the large values of
$\vert g\vert$ and $\alpha$ \cite{khodaparast}. We have considered the 
following parameters for the InSb quantum dot: $m^*/m_0=0.014,
\epsilon=17.88, g=-40$ and $\hbar\omega_0=7.5$ meV. The energy levels for
InSb quantum dots containing 2 -- 4 interacting electrons are plotted in
Figs.~\ref{fig:E2_InSb}, \ref{fig:E3_InSb}, \ref{fig:E4_InSb} and
for various values of the SO coupling strength $\alpha$. The corresponding
optical absorption spectra are presented in Figs.~\ref{fig:OA2_InSb},
\ref{fig:OA3_InSb}, and \ref{fig:OA4_InSb}. As compared to the spectra of 
InAs dots a clear difference is the almost total absence of anticrossings 
and discontinuities. This is partly due to the very large Zeeman coupling 
which practically nullifies the SO interaction at the coupling strengths
$\alpha$ we are concerned with. Another reason is the large kinetic energies 
due to the very small electron effective mass. Because the strength of the 
Coulomb interaction is somewhat smaller than in InAs ($\epsilon_{\rm InSb}
> \epsilon_{\rm InAs}$) correlations caused by the mutual electronic 
interactions are effectively much smaller in InSb than in InAs. For the
exploration of the SO coupling via absorption spectroscopy, InSb quantum 
dots do not seem to be a very promising system.

\begin{figure}[t]
\begin{center}
 \includegraphics[width=.45\textwidth]{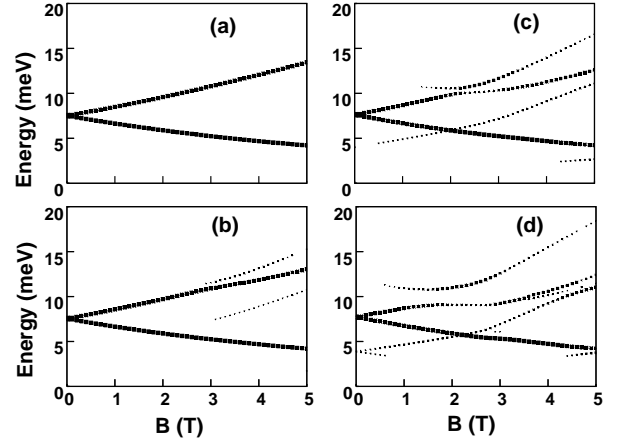}
\protect\caption{Same as in Fig.~\protect\ref{fig:OA2_InAs}, but for
a two-electron GaAs quantum dot.
}\label{fig:OA2_GaAs}
\end{center}
\end{figure}

\subsection{GaAs quantum dots}

\begin{figure}[t]
\begin{center}
 \includegraphics[angle=-90, width=.45\textwidth]{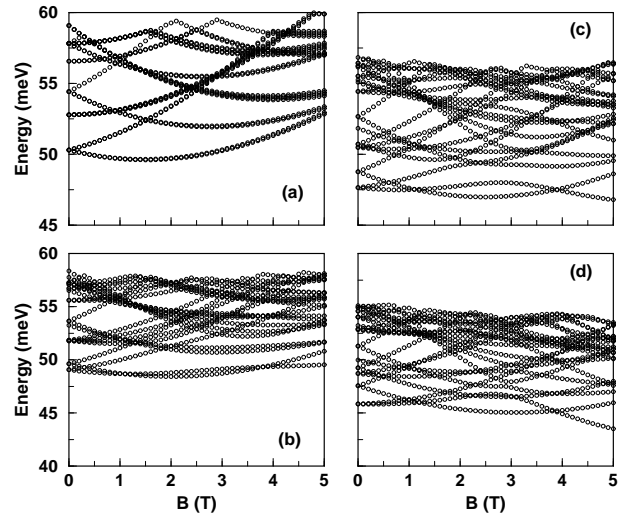}
\protect\caption{Same as in \protect\ref{fig:E2_InAs}
but for a three-electron GaAs quantum dot.
}\label{fig:E3_GaAs}
\end{center}
\end{figure}

\begin{figure}[b]
\begin{center}
 \includegraphics[width=.45\textwidth]{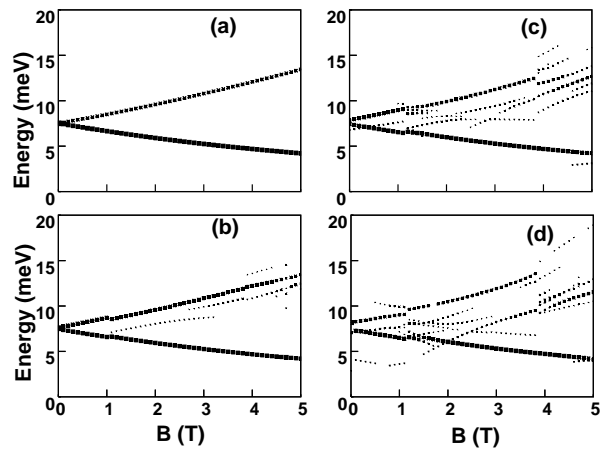}
\protect\caption{Same as in Fig.~\protect\ref{fig:OA2_InAs}, but for
a three-electron GaAs quantum dot.
}\label{fig:OA3_GaAs}
\end{center}
\end{figure}

\begin{figure}[b]
\begin{center}
 \includegraphics[angle=-90, width=.45\textwidth]{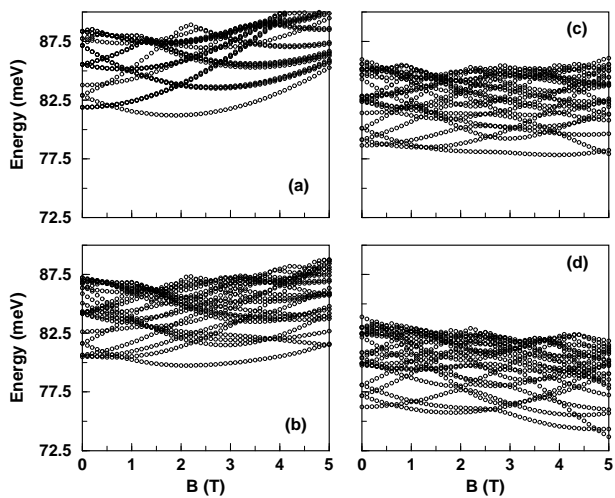}
\protect\caption{Same as in \protect\ref{fig:E2_InAs}
but for a four-electron GaAs quantum dot.
}\label{fig:E4_GaAs}
\end{center}
\end{figure}

\begin{figure}[t]
\begin{center}
 \includegraphics[width=.45\textwidth]{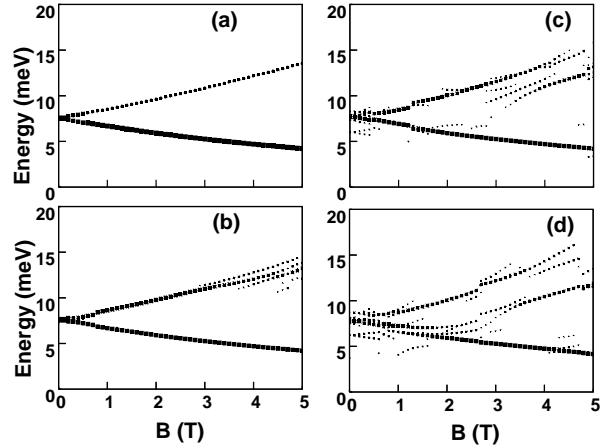}
\protect\caption{Same as in Fig.~\protect\ref{fig:OA2_InAs}, but for
a four-electron GaAs quantum dot.
}\label{fig:OA4_GaAs}
\end{center}
\end{figure}

The results for GaAs quantum dots, ones that are most intensely explored in 
the absence of SO coupling, are presented here primarily as an academic interest. 
The parameters that we have used here are: $m^*/m_0=0.063, \epsilon=12.9, 
g=-0.44$ and $\hbar\omega_0=7.5$ meV. Clearly, the very low value of the
$\vert g \vert$-factor perhaps makes the GaAs QDs unsuitable for any 
observable effect due to the SO coupling. Interestingly, however, among 
all the three types of QDs studied here for optical absorptions, GaAs QDs 
show the most spectracular effects for large values of $\alpha$. The energy 
levels for GaAs quantum dots containing 2 -- 4 interacting electrons are 
plotted in Figs.~\ref{fig:E2_GaAs}, \ref{fig:E3_GaAs}, and \ref{fig:E4_GaAs}
for various values of the SO coupling strength $\alpha$. The corresponding
optical absorption spectra are presented in Figs.~\ref{fig:OA2_GaAs},
\ref{fig:OA3_GaAs}, and \ref{fig:OA4_GaAs} respectively. As mentioned above, 
the only observed value of $\alpha$ for GaAs QD reported as yet is $\alpha\sim6$ 
meV.nm \cite{koenemann}. Reversing the arguments presented in the previous 
subsection, i.e. in GaAs a very small Zeeman coupling and a rather large effective 
electron mass but practically equal strength of Coulomb interaction help us
understand why the absorption spectra of our GaAs dots exhibit a remarkably rich 
structure as opposed to those of the InSb and InAs dots. Finding an appropriate 
set up to generate a large $\alpha$ for GaAs quantum dot would be a major
(but worthwhile) experimental endeavor.

\section{A brief review of earlier theoretical works}

In this section, we present a critical review of earlier theoretical reports
on how the Bychkov-Rashba spin-orbit coupling in parabolic quantum dots were
treated \cite{kuan,governale,vam_PRB,vam_JAP,destefani,debald,koenemann,quasi_exact,%
tsitsishvili,manuel_041,manuel_042,manuel_043,manuel_02,lucignano,cremers,bellucci,fransson}.
There are quasi-exact solutions available for electrons confined in a parabolic
quantum dot in the presence of the SO interaction, but without the inter-electron
interaction \cite{quasi_exact}, and exact analytical results are also reported
in the case of a circular quantum dot with hard walls \cite{tsitsishvili},
again for a non-interacting system, but with the Bychkov-Rashba spin-orbit
interaction included. However, for the realistic systems of parabolic quantum 
dots with interacting electrons these methods are prohibitively complicated
and evaluation of the energy spectrum can only be done numerically.
Among the theoretical papers dealing with the SO interaction in quantum dots
discussed below, Kuan et al. \cite{kuan} presented the best treatment of the
single-electron states. They looked at the energy levels of parabolically 
confined quantum dots with Bychkov-Rashba SO coupling and in the presence of
zero and nonzero magnetic fields. They solved the single-particle equation
correctly by expanding the solution spinors in terms of the eigenfunctions
of QDs without the SO interaction, i.e. the Laguerre functions. We have used
a similar approach to construct the basis states for our multi-electron QDs
[Sects.~II, III].

Voskoboynikov et al. \cite{vam_PRB} studied the effect of SO interaction on
the energy spectrum of cylindrical semiconductor QDs in an externally applied
magnetic field. They considered the Bychkov-Rashba SO coupling due to the 
parabolic confinement, i.e. an in-plane field conserving orbital and spin
angular momenta. As a consequence, the SO coupling has no qualitative effects on, 
for example, the absorption spectra. The electron-electron interaction was not
included in this scheme. In Ref.~\cite{vam_JAP}, they studied the magnetization
and magnetic susceptibility in few-electron parabolic QDs with SO coupling. As
in their earlier paper, they handled the SO term only due to the parabolic
confinement and therefore the single-particle Hamiltonian is diagonal in spin
space. They neglected the mutual electronic Coulomb interaction.

Governale \cite{governale} investigated the effects of SO coupling on the
addition energy and on the spin properties of few-electron QDs in the absence
of the external magnetic field. He introduced the SO coupling into single-particle
states perturbatively but also compared the resulting energies to the ones obtained
by numerical diagonalization technique. At the small SO coupling strength that
he considered, the perturbation approach seems to be valid. Electron correlations
were handled by using the spin-density functional approach. Cremers et al. \cite{cremers}
studied conductance and its fluctuations in the presence of SO interaction,
Zeeman coupling, externally applied magnetic field, in a (single-electron) QD.
They solved the single-particle equation applying an approximate unitary
transformation which in leading order takes the Hamiltonian to a diagonal form in
spin space.

Valin-Rodriguez \cite{manuel_041} considered a single electron in a parabolic QD with
Bychkov-Rashba SO coupling. He performed a unitary transformation to transform the
Hamiltonian in spin space to a diagonal form up to second order in SO and Zeeman
coupling parameters. He showed that the effective SO interaction is influenced by
the interplay between Zeeman and SO couplings. In Ref.~\cite{manuel_042}, Valin-Rodriguez 
et al. introduced a spatially modulated (in radial direction) Bychkov-Rashba coupling
in single-electron (disk) QDs. They solved the two-component spinor equation and numerically
evaluated the spin density. They concluded that it is possible to confine electrons
spatially with appropriate structural modulation. These authors also investigated
the SO couplings in deformed parabolic quantum dots \cite{manuel_043}. They solved the 
single-particle equations using the approximate unitary transformations mentioned above. 
They were interested in the effects of spatial deformations to the spin splitting oscillations.
They estimated the Coulomb interaction contribution using the time-dependent local-%
spin-density approximation \cite{manuel_02}.

Lucignano et al. \cite{lucignano} studied the few-electron QDs including the mutual
electron-electron interaction and under the influence of an externally applied
magnetic field. They applied an exact diagonalization method (but with a rather
restricted basis: 28 single-particle states deducing from their earlier
paper \cite{jouault}. They particularly looked at the possibility to use the
SO coupling to control the excitations under the magnetic fields which polarize
the ground state, i.e., close to the final {\it single-triplet} transition. The 
SO coupling is included in the many-electron Hamiltonian, but not in the basis
states. They evaluated the dipole matrix elements for absorption from the
ground state to the lowest dipole-allowed excited state. They claimed that
there is an increase in intensity close to the transition to the fully polarized
ground state.

Destefani et al. \cite{destefani} reported numerical results for energy 
levels and spin polarizations for one and two-electron parabolic QDs 
under a magnetic field and the SO coupling. For the two-electron system, 
Coulomb interaction is also included. They do not construct the correct 
single-electron states in the two-electron QD, the SO coupling is, in fact,
taken into account only in the many-electron Hamiltonian. Debald et al. 
\cite{debald} studied oscillations in few-electron parabolic QD in a magnetic 
field, between states where the degeneracy is lifted by the SO coupling, i.e. 
at the level repulsion points. The Coulomb interaction was taken into account 
only approximately, because the many-body effects were claimed to play 
only a minor role in the very small magnetic field considered in that 
work. K\"onemann et al. \cite{koenemann} considered the SO coupling in single 
electron QDs. They showed that there is an anisotropy between spin 
splittings due to magnetic fields parallel and perpendicular to the
dot. The anisotropy was shown to be proportional to the strength of the SO 
coupling. Bellucci and Onorato \cite{bellucci} studied the influence of SO 
coupling on the charge and spin polarization in a vertical disk-shaped QD 
under a strong perpendicular magnetic field. They treated the SO coupling 
perturbatively (upto second order). They handled the Coulomb interaction 
within the Hartree-Fock approach. They studied the energy splittings due 
to the SO coupling. Finally, Fransson et al. \cite{fransson} studied transport 
through QDs with spin dependent couplings to the contacts. They evaluated 
the QD energy levels using a (first principles) density functional theory. 
They calculated the transport properties of (a) non-interacting electrons 
taking into account the few levels closest to the Fermi level, and (b) 
interacting electrons using an approximate Hamiltonian with the levels 
closest to the Fermi level. We would like to note here that, in the light 
of all these theoretical approaches, our method of including the SO coupling 
for interacting electrons in a parabolic QD seems to be the most accurate one. 
Although, given the fact that our approach involves extensive numerical 
computations, some of the approaches discussed above, such as the one by 
Lucignano et al. \cite{lucignano}, seem to be very promising.

\section{Conclusions}

In conclusion, we have studied the energy levels and optical-absorption
spectra for parabolic quantum dots containing upto four interacting
electrons, in the presence of spin-orbit coupling and under the
influence of an externally applied, perpendicular magnetic field.
We have presented a very accurate numerical scheme to evaluate these quantities.
We have presented results for the Fock-Darwin spectra in the presence of
SO coupling for quantum dots made out of three different semiconductor systems,
InAs, InSb, and GaAs. The effects of SO coupling on the single-electron
spectra are primarily to lift the degeneracy at $B=0$, rearrangement of
some of the energy levels at small magnetic fields, and level repulsions
at high fields. These are explained as due to mixing of different spinor
states for increasing strength of SO coupling. As a consequence, the
corresponding absorption spectra reveal anticrossing structures in the two
main lines $(\alpha=0)$ of the spectra. For the interacting many-electron
systems we observed the appearence of discontinuities, anticrossings, and
new modes that appear in conjunction with the two main absorption lines. 
These additional features arise entirely due to the SO coupling and are a
consequence of level crossings and level repulsions in the energy spectra.
An intricate interplay between the SO coupling and the Zeeman energies are
shown to be responsible for these new features seen in the energy spectra.
Our accurate results for the low-lying energy levels for the SO coupled
QDs can also be measured, in principle, by transport \cite{rolf_review}
or capacitance \cite{ashoori} spectroscopy, which have been successfully
employed earlier to map out the energy spectra of parabolic quantum dots.
Optical absorption spectra for all three types of quantum dots containing
a few interacting electrons that are studied here
show a common feature: new modes appear, mostly near the upper main branch 
of the spectra around 2 tesla that become stronger with increasing $\alpha$.
Among the three types of systems considered here, optical signatures of the 
SO interaction is found to be the strongest in the absorption spectra of a 
GaAs quantum dot, but only at very large values of the SO coupling strength, 
and appears to be the weakest for the InSb quantum dots. Experimental 
observation of these new optical modes that appear solely due to the presence of 
SO coupling would be very exciting because that would be a major step forward 
in our quest to manipulate the spin dynamics in nanostructured systems via 
the SO coupling.

Our future works along this line will be to explore coupled QDs, or QD molecules
\cite{austing}. Our primary goal will be to generate accurate results for energy
levels and optical absorption spectra for coupled (laterally \cite{laterally} 
or vertically \cite{vertically}) quantum dots with spin-orbit interaction.
In addition to being important for fundamental studies, these results would
be interesting from the point of view of quantum computations \cite{coupled_bits}
as well. It is now well recognized that semiconductor quantum dots have the
potential to become the building blocks for solid state quantum computation.
Quantum states of single, double and even triple coupled \cite{triple_coupled}
quantum dots have been explored for this purpose. In a quantum computer, 
information is stored in a two-level system. Hence a promising candidate system 
to realize the quantum bits, the fundamental unit of information in a quantum 
computer, is a quantum dot where the single-electron states can be used for 
that purpose. In a magnetic field, the Zeeman splitting of electron spin 
can provide a two-level system (for an odd number of electrons). Alternatively, 
the spatial wave function of a single-electron state in a double quantum 
dot (allowing for electron tunneling between the two dots) can also represent 
a two-level system. Spin-orbit coupling in the coupled quantum dot systems 
could perhaps be used to perform quantum computation (using the spin rotation, 
for example). Accurate results for the energy levels of the coupled-dot system 
might be beneficial in that direction of research. The effects of SO coupling
on the energy levels and absorption spectra for a more complex system
such as coupled QDs are, however, important and interesting in their own ways.
These would be the subjects of our future publications.

\section{ACKNOWLEDGMENTS}
The work of T.C. has been supported by the Canada Research Chair 
Program and the Canadian Foundation for Innovation (CFI) Grant.
We wish to thank Dr. Alex Voskoboynikov for a critical reading of the
manuscript.

\appendix*

\section{Diagonalization of monster matrices}

We have mentioned in Sect.~III about the challenging task of construction 
of the Hamiltonian matrix and its numerical diagonalization. Here we present 
a brief discussion about the numerical method that we believe the most 
accurate (and appropriate) for that task. While in principle, evaluation of 
Eq.~(\ref{coulelsum}) is straightforward it turns out to be numerically 
highly unstable, primarily due to the expansion (\ref{gnmexp}) extending 
to Laguerre polynomials of large degree and large angular momenta which in 
turn, leads to large terms of alternating sign \cite{stone}. A remedy
for this is to employ multiple precision arithmetics such as,
for example implemented in the Gnu arbitrary precision GMP
library. However, if we apply multiple precision arithmetic
directly into the sixfold summation (\ref{coulelsum}) the time consumed
to evaluate these becomes insurmountable. To circumvent this obstacle
we note that many terms in the sums actually depend on very
few parameters, the range of these parameters is restricted
and the same functional forms repeat themselves.
Thus a natural solution is to tabulate these forms and the subsums.
In our Coulomb matrix element code we used the tabulated functions
\begin{eqnarray*}
D(i) &=& i! \\
F(n,\ell,\kappa)&=&\frac{(n+\ell)!}{(n-\kappa)!(\ell+\kappa)!}\\
&=&\frac{D(n+\ell)}{D(n-\kappa)D(\ell+\kappa)} \\
G(n,\ell,\kappa)&=&\frac{(n+\ell)!}{\kappa!(n-\kappa)!(\ell+\kappa)!}
=\frac{F(n+\ell)}{D(\kappa)} \\
H(s)&=&\frac{\Gamma(s+\frac12)}{2^{s+1}} \\
I(q_1,q_2,\ell)&=&
D(q_1)D(q_2)\sum_{s=0}^{q_1}(-1)^sG(q_1,\ell,s) \nonumber \\
&&\times\sum_{t=0}^{q_2}(-1)^tG(q_2,\ell,t)H(t+s+\ell).
\end{eqnarray*}
Now the summations $\Sigma$ in the expression (\ref{coulelsum}) can
be written as
\begin{eqnarray}
\Sigma&=&\sum_{\kappa_1=0}^{n_1}(-1)^{\kappa_1}G(n_1,\ell_1,\kappa_1)
\sum_{\kappa_2=0}^{n_2}(-1)^{\kappa_2}G(n_2,\ell_2,\kappa_2)
\nonumber \\
&&\times
\sum_{\kappa_3=0}^{n_3}(-1)^{\kappa_3}G(n_3,\ell_3,\kappa_3)
\sum_{\kappa_4=0}^{n_4}(-1)^{\kappa_4}G(n_4,\ell_4,\kappa_4)
\nonumber \\
&&\times
I(\kappa_1+\kappa_4+\tfrac12(\ell_1+\ell_4-k),
\kappa_2+\kappa_3 \nonumber \\
&&+\tfrac12(\ell_2+\ell_3-k)).
\label{couleltabsum}
\end{eqnarray}
Although the summation here is still fourfold it is nevertheless
several orders of magnitude faster than the original
one (\ref{coulelsum}) and as such fast enough for our purposes.

Under the influence of the SO coupling the total spin $S$ of
our many-electron system is not a conserved quantity. As a
consequence of this we cannot fix the total $S_z$ of the
many-body basis. This degree of freedom tends to make the
number of non-interacting many-body states of the basis very
large even for a small number of electrons, and even if the
conservation of the total angular momentum $J_z=L+S_z$ is
taken into account. For example,
to achieve a convergence for a four electron system in
the parabolic confinement with harmonic potential ($\hbar\omega_0$)
of few meV the size of the basis must be {\it of the order of a million}.
Furthermore, since we want to study  properties of the eigenstates,
such as polarization and dipole matrix elements between states,
we also need the relevant eigenvectors. Clearly the sheer size
of the matrix prohibits a full diagonalization and we have to
resort to an iterative scheme aimed to search a given number of
energetically lowest eigenvectors. Of course the algorithm
should be fast and hopefully also robust.

The algorithm proposed by Davidson and Liu (DL) to evaluate the
eigenvalues and eigenvectors of ``monster matrices'' \cite{davidson} 
seem to fit our criteria. Like any other iterative method it transforms 
the diagonalization of a matrix $A$ to minimization of
the Rayleigh quotient
\begin{equation}
\lambda=\frac{x^{\scriptsize T} Ax}{x^{\scriptsize T}x},
\label{minprinc}
\end{equation}
where $x$ represents the column vector of the coefficients in
the superposition of the basis states. Also, as in many other methods, 
the only operations involving the matrix $A$ are vector multiplications. 
This allows us to exploit fully the sparseness of $A$, i.e. we have to store
only the non-zero elements.

The key idea behind practically any iterative method subjecting
the matrix only to multiplication is to search the minimum
of the quotient (\ref{minprinc}) in a (very small) subspace
and to update this subspace in each iteration step. How this
updating is performed depends on the method. For example, in
the common conjugate gradient method the subspace is two-%
dimensional and spanned by the gradient $g$ at current position $x$
and a vector $s$ conjugate to it with respect to $A$ 
(i.e. $s^{\scriptsize T} A g=0$).

In the DL method the dimension of the search space varies from step to 
step. Suppose that at a given step our search space $\cal S$ is spanned by the
orthonormal vectors $s_1,\ldots,s_K$ (the dimension $K$ must, of course,
be greater than the number of required eigenstates). Finding the minimum 
of (\ref{minprinc}) in this subspace corresponds to the diagonalization 
of the $K\times K$ matrix ${\cal S}^{\scriptsize T}A{\cal S}$ (we take $\cal S$ 
to represent also the matrix with columns $s_k$). As a result we get $K$ eigenvalues 
$\mu_k$ and $K$ eigenvectors $z_k$, each of dimension $K$. The expanded vectors
$$ y_k=\sum_{i=1}^Kz_{k,i}s_i $$
will approximate the eigenvectors and the quantities $\mu_k$
the eigenvalues we are seeking for. The next task in the iteration
step is to update the subspace $\cal S$. For that purpose we pick up
a certain number (a parameter depending for example on the
size of the computer memory) of the residuals
$$ r_k=(A-\mu_k)y_k $$
with largest norms. The selected residuals are orthonormalized with respect 
to the space $\cal S$ and then appended to it. So, in each step the dimension of 
the search space and the size required to store it increases and we may 
eventually exhaust all the memory. At this point we compress the space $\cal S$ to 
its bare minimum comprising only as many vectors as we are required to find. 
These vectors are selected from the set $\{y_k\}$ and are the ones with smallest 
$\mu_k$.

\end{document}